\documentclass[12pt]{article}

\usepackage{amsmath}
\usepackage{amsfonts}
\usepackage{amssymb}
\usepackage{graphics,graphicx,tikz}
\usepackage{soul}
\usepackage{tensor}
\usepackage{makecell}
\usetikzlibrary{patterns}
\numberwithin{equation}{section} 
\usepackage[linktocpage]{hyperref}
\hypersetup{
	colorlinks=true,
	linkcolor=blue,
	filecolor=magenta,      
	urlcolor=blue,
	citecolor=blue
}
\def\beq{\begin{equation}} 
\def\eeq{\end{equation}}
\usepackage{color}

\usepackage[noadjust]{cite}

\begin{document}
\begin{titlepage}
\begin{flushright}
\end{flushright}
\vskip 1.0cm
\begin{center}
{\Large \bf Revisiting gravitational angular momentum and mass dipole losses in the eikonal framework}
\vskip 1.0cm {Carlo Heissenberg$^{\dagger}$,
	Rodolfo Russo$^{\dagger}$\footnote{\texttt{r.russo@qmul.ac.uk}}} \\[0.7cm]
{\it \small$^\dagger$School of Mathematical Sciences,
	Queen Mary University of London, \\
	Mile End Road, London, E1 4NS, United Kingdom.}
\end{center}

\vspace{5pt}

\begin{abstract}
We review the description of classical gravitational scatterings of two compact objects by means of the eikonal framework. This encodes via scattering amplitudes both the motion of the bodies and the gravitational-wave signals that such systems produce. As an application, we combine the next-to-leading post-Minkowskian (PM) waveform derived in the post-Newtonian (PN) limit with the 4PM static loss due to the linear memory effect to reproduce known results for the  
total angular mometum loss in the center-of-mass frame up to $\mathcal O(G^4)$ and 2.5PN order. We also provide similar expressions for the change in the system's mass dipole, discussing the subtleties related to its sensitivity to the Coulombic components of the field and to the nonlinear memory effect.
\end{abstract}
\end{titlepage}

\tableofcontents

\section{Introduction}
\label{sec:intro}

In gravitational theories, the standard perturbative approach to scattering amplitudes breaks down at transplanckian energies $E\gg \sqrt{\hbar/G}$, as the effective gravitational coupling $GE^2/\hbar$ becomes large. It is not surprising, then, that the problem of head-on collisions at high energies is very challenging and, at the same time, very interesting, since it is closely related to the issue of black-hole formation and unitarity. The regime where the incoming states are well separated is instead much more tractable. 

In this regime, which corresponds to the post-Minkowskian (PM) limit, it is possible to rearrange the standard perturbation theory to resum the large contributions due to the couplings between the highly energetic states and the gravitons.
An efficient way to implement this idea goes under the name of \emph{gravitational eikonal} and has been studied in great detail starting from the eighties~\cite{'tHooft:1987rb,Amati:1987wq}. At the time, the focus was more on conceptual issues and on string theory, while more recently, following~\cite{Neill:2013wsa,Damour:2016gwp,Damour:2017zjx,Bjerrum-Bohr:2018xdl,Cheung:2018wkq}, this approach has been applied to the study of black hole encounters, see~\cite{DiVecchia:2023frv} and references therein. The basic idea is that the dominant contributions to the $S$-matrix due to the large energy $E$ take a simpler form after performing a Fourier transform to impact parameter space, where they exponentiate to define a classical quantity: the gravitational eikonal. 

Here we will use this framework to describe the scattering of two massive scalar particles minimally coupled to gravity in four spacetime dimensions. 
As mentioned, we work in the regime where the impact parameter is much larger than the effective size $GE$ and the goal is to provide a quantitative characterization of the final state. The  momentum of the massive particles changes compared to the initial state because of the mutual gravitational interaction, while at the same time radiation is produced in the form of  gravitational waves. 
Thus, it is convenient to describe the final state as an operator acting on the Fock space of the graviton modes. This eikonal operator \cite{Damgaard:2021ipf,Cristofoli:2021jas,DiVecchia:2022piu,DiVecchia:2023frv} involves two main ingredients: the elastic $2 \to 2$ amplitude, with external massive states representing the black holes, and the inelastic $2\to3$ amplitude with the emission of one graviton. 
We will use these inputs up to 3PM, which means at  next-to-leading (NLO) order for the 5-point amplitude and at NNLO for the 4-point amplitude. 

There is by now a vast literature on these amplitudes, their classical limit and the observables that can be obtained from them. 
The deflection of the massive objects, also referred to as classical impulse, was analyzed within different approaches
 up to 3PM in~\cite{Bern:2019nnu,KoemansCollado:2019ggb,Bern:2019crd,Cristofoli:2020uzm,Parra-Martinez:2020dzs,DiVecchia:2020ymx,DiVecchia:2021ndb,Herrmann:2021lqe,DiVecchia:2021bdo,Herrmann:2021tct,Bjerrum-Bohr:2021vuf,Heissenberg:2021tzo,Bjerrum-Bohr:2021din,Damgaard:2021ipf,Brandhuber:2021eyq}, while the state of the art for this observable goes beyond 3PM, with a complete analysis at 4PM~\cite{Bern:2021dqo,Bern:2021yeh,Dlapa:2022lmu,Dlapa:2023hsl,Damgaard:2023ttc} and partial results at 5PM~\cite{Driesse:2024xad,Bern:2024adl}.
 The inelastic $2\to3$ amplitude and its relation with the classical gravitational waveform was discussed instead in~\cite{Goldberger:2016iau,Luna:2017dtq,Mogull:2020sak,Jakobsen:2021smu,Mougiakakos:2021ckm} at LO and~\cite{Brandhuber:2023hhy,Herderschee:2023fxh,Elkhidir:2023dco,Georgoudis:2023eke,Caron-Huot:2023vxl,Georgoudis:2023lgf,Georgoudis:2023ozp} at NLO. 
 When it comes to extracting observables from amplitudes, the eikonal operator framework can be thought of as a  reorganization of the Kosower--Maybee--O'Connell \cite{Kosower:2018adc} strategy, whereby classical gravitational observables are given by expectation values of  operators in the final states obtained by the action of the $S$-matrix \cite{Damgaard:2023vnx}. A closely-related line of development is based on efficiently solving the classical problem via worldline methods \cite{Kalin:2020mvi,Kalin:2020fhe,Mogull:2020sak,Dlapa:2021npj,Jakobsen:2022psy,Kalin:2022hph,Dlapa:2022lmu,Dlapa:2023hsl,Driesse:2024xad}.

Our main objective in this work is to use the eikonal operator to obtain explicit expressions for the full Lorentz tensor 
\begin{equation}\label{eq:Jsum}
	J_{\alpha\beta} = \boldsymbol{J}_{\alpha\beta} + \mathcal{J}_{\alpha\beta} +\delta \mathcal{J}_{\alpha\beta} 
\end{equation}
describing in a covariant way the angular momentum and mass dipole moment lost by the binary to the gravitational field. 
We will discuss more in detail each term in \eqref{eq:Jsum} below, but the general convention is to use boldface symbols for quantities that are entirely determined by  gravitons with non-zero frequency and calligraphic symbols for contributions that depend on the static gravitational field and in particular on the ``zero-frequency gravitons''.
The covariant angular momentum is closely connected to various subtleties related to low-frequency gravitational modes, as already evident in the full PM results derived in~\cite{Damour:2020tta,Bini:2022wrq} at $O(G^2)$ and in~\cite{Manohar:2022dea,DiVecchia:2022piu} at  $O(G^3)$.
In particular, while dissipation of energy-momentum only takes place via the emission of dynamical gravitational waves implying $P_\alpha = \boldsymbol{P}_\alpha$, the two-body system can lose angular momentum both via the emission of physical radiation modes, gravitational waves, and due to the interaction with the static gravitational field~\cite{Damour:2020tta,Manohar:2022dea,DiVecchia:2022owy}.  The latter effect is also intimately tied to BMS supertranslations, as emphasized by \cite{Veneziano:2022zwh}, as well as to the memory effect \cite{Zeldovich:1974gvh,Christodoulou:1991cr,Wiseman:1991ss,Thorne:1992sdb} and the Weinberg soft graviton theorem \cite{Manohar:2022dea,DiVecchia:2022owy}, and is thus closely related to the core ``infrared triangle'' characterizing  the asymptotic properties of gravity \cite{Strominger:2014pwa,He:2014laa,Strominger:2017zoo}.
While several alternative formulas for the angular momentum loss have been proposed \cite{Compere:2019gft,Chen:2021szm,Chen:2021kug,Javadinezhad:2022ldc,Mao:2023evc,Riva:2023xxm,Javadinezhad:2023mtp,Mao:2024ryq}, both kinds of contributions are taken into account by the manifestly covariant formula for $J_{\alpha\beta}$ of Refs.~\cite{Manohar:2022dea,DiVecchia:2022owy}, which we take as our starting point, being particularly natural in the amplitude approach. 
 
We discuss how these two contributions are encoded in the approach of the eikonal operator. For the radiative part $\boldsymbol{J}_{\alpha\beta}$, we review how to connect the formulas used in the covariant amplitude approach to the expressions used in the post-Newtonian (PN) literature in terms of the transverse-traceless (TT) waveform \cite{Maggiore:2007ulw,Blanchet:2013haa,Damour:2020tta} or equivalently of its multipole decomposition~\cite{Blanchet:2013haa,Compere:2019gft,Bini:2021gat,Bini:2022enm}. Then, we include in the eikonal operator the information encoded by the NLO waveform, which was recently obtained in the PN limit from the small-velocity limit of the PM answer \cite{Bini:2023fiz,Georgoudis:2024pdz,Bini:2024rsy}. From this, we obtain explicit expressions for $\boldsymbol{J}_{\alpha\beta}$ in the small-velocity expansion, both for the spatial part $\boldsymbol{J}_{ij}$, the angular momentum proper, and for the mixed space-time components $\boldsymbol{J}_{i0}$, the mass dipole, in the center-of-mass frame.
 As we shall see, while $\boldsymbol{J}_{ij}$, is perfectly well defined and unambiguous, the $\boldsymbol{J}_{i0}$ components retain a non-trivial contribution coming from the tail part of the waveform~\cite{Blanchet:1987wq,Blanchet:1992br,Blanchet:1993ng} and inherit a dependence on the arbitrary scale related to shifts of the retarded time induced by the tail effect \cite{Goldberger:2009qd,Porto:2012as,Brandhuber:2023hhy,Herderschee:2023fxh,Elkhidir:2023dco,Georgoudis:2023eke}. This holds both in the eikonal and in the standard PN approach.
 Following \cite{Blanchet:2013haa,Blanchet:2018yqa,Compere:2019gft} however, in the center-of-mass frame one can subtract the drift due to recoil by defining $\boldsymbol{M}_i = \boldsymbol{J}_{i0}-\int t  \dot{P}_i dt$, and this quantity is then well defined and independent of the aforementioned arbitrary scale.
  Note that 
   $\dot{J}_{i0} = - \dot{L}_{i0} = \dot{Z}_i$,
  where $L_{\alpha\beta}$ is the mechanical angular momentum tensor and, following \cite{Blanchet:2018yqa}, $Z_i$ represents the intial position of the center of mass  of the system multiplied by the total energy. We recall that, in the absence of radiation, the three components $Z_i$ represent the conserved quantities associated to the invariance of the theory under Lorentz boosts, in the same way as the three components of the angular momentum are conserved due to invariance under spatial rotations.
 
 Then we turn our attention to the zero-frequency contributions, which were first included in the eikonal setup in \cite{DiVecchia:2022owy} (see also \cite{DiVecchia:2022nna,DiVecchia:2022piu,Cristofoli:2022phh,DiVecchia:2023frv}) by taking into account the interference effects between the static field and the soft waveform determined by the \emph{linear} memory effect \cite{Zeldovich:1974gvh,Weinberg:1964ew,Weinberg:1965nx}. This strategy, already outlined in \cite{Manohar:2022dea}, allowed \cite{DiVecchia:2022owy} to obtain a formally all-order expression for this type of effect, $\mathcal{J}_{\alpha\beta}$, see Eq.~(3.30) of that reference. Contributions arising from the \emph{nonlinear} memory part of the soft waveform~\cite{Christodoulou:1991cr,Wiseman:1991ss,Thorne:1992sdb}, which we denote by $\delta\mathcal{J}_{\alpha\beta}$, are instead expected to enter at $\mathcal{O}(G^4)$ by power counting and had been neglected in \cite{DiVecchia:2022owy}. Building on the approach of that reference, we derive here a general expression for $\delta\mathcal{J}_{\alpha\beta}$ and discuss its properties.
  
First, we show that the spatial components  $\delta\mathcal{J}_{ij}$ are unambiguous at least up to and including $\mathcal{O}(G^6)$.
Second, we find that $\delta\mathcal{J}_{ij}$ does not include any $\mathcal{O}(G^4)$ contribution, while it is nontrivial at order ${\cal O}(G^5)$.
This means that the full zero-frequency contribution at $\mathcal{O}(G^4)$ is captured by $\mathcal{J}_{ij}$, for which we provide an explicit PM expression for generic velocities at order $\mathcal{O}(G^4)$. Combining this with the radiative part, obtained following the above strategy in the small-velocity limit, we verify that $\boldsymbol{J}_{ij}+\mathcal{J}_{ij}$ perfectly agrees up to $\mathcal{O}(G^4)$ and 2.5PN order with results obtained in Refs.~\cite{Bini:2021gat,Bini:2022enm} for this observable.
This nontrivial cross-check, whereby amplitude-based results match previous PN expressions obtained from purely classical methods, supports the picture presented above in terms of radiative and static contributions, and in particular the idea that the supertranslation of Ref.~\cite{Veneziano:2022zwh} is exact in $G$.
Finally, the nonlinear effect enters in a nontrivial way the mass dipole components $\delta\mathcal{J}_{i0}$ already at $\mathcal{O}(G^4)$ and induces the same sensitivity to choice of the origin of retarded time mentioned above for the radiative contribution $\boldsymbol{J}_{i0}$. 
While there it entered due to the infrared divergence associated to the tail effect, here the time-shift ambiguity arises due to a collinear singularity.
For completeness, we include the PM expression for $\mathcal{J}_{i0}$ up to $\mathcal{O}(G^4)$ and the result for $\delta\mathcal{J}_{i0}$ at $\mathcal{O}(G^4)$ in the small-velocity limit.

Another interesting feature appearing in the analysis of the static mass dipole components $\mathcal{J}_{i0}+\delta{\mathcal{J}}_{i0}$ is their sensitivity to the Coulombic field of the binary system. As noticed in~\cite{Riva:2023xxm}, this is in fact an essential feature associated to the Lorentz covariance of $J_{\alpha\beta}$. As we will see, the effect of the Coulombic field can be automatically taken into account 
by using all modes in the de Donder gauge rather than restricting the static part of the eikonal operator to the TT part (cf.~the discussions in \cite{Ashtekar:2017ydh,Ashtekar:2017wgq,Bonga:2018gzr}). 
Another subtlety associated to a rearrangement between field and mechanical contributions is the so-called scoot, see  \cite{Gralla:2021qaf,Gralla:2021eoi,Bhardwaj:2022hip,Bini:2022wrq,Alessio:2024crv,Gralla:2024wzr}.
	
We also derive the analogous expressions involving the TT projection of the field, $\mathcal{J}^{\text{TT}}_{\alpha\beta}+\delta{\mathcal{J}}^{\text{TT}}_{\alpha\beta}$, which instead excludes the Coulombic field contributions, and show that, while the angular momentum components coincide with the above ones, $\mathcal{J}_{ij}=\mathcal{J}^\text{TT}_{ij}$, $\delta\mathcal{J}_{ij}=\delta\mathcal{J}^\text{TT}_{ij}$, the mass dipole components are different and in fact are free of collinear singularities and independent of the above mentioned time shift. This comes at the price of losing Lorentz covariance, since the resulting formula depends on the form of the reference vector entering the TT projection. We are thus led to define $\mathcal{M}_{i}=\mathcal{J}^{\text{TT}}_{i0}$  (note that static modes do not contribute to the radiated energy-momentum) and similarly $\delta\mathcal{M}_{i}=\delta\mathcal{J}^{\text{TT}}_{i0}$. Once again, we present explicit PM formulas for $\mathcal{M}_i$ up to $\mathcal{O}(G^4)$, and fall back to the PN expansion for $\delta\mathcal{M}_i$. To summarize, 
\begin{equation}
	J_{ij} 
	= 
	\boldsymbol{J}_{ij}
	+
	\mathcal{J}_{ij}
	+
	\delta\mathcal{J}_{ij}\,,
	\qquad
	M_i 
	=
	\boldsymbol{M}_{i} 
	+
	\mathcal{M}_{i} 
	+
	\delta\mathcal{M}_{i} 
\end{equation}
provide unambiguous expressions for the total variation of the ten charges associated to Poincar\'e invariance in the center-of-mass frame. See Table~\ref{table:properties} for a visual guide to the properties of the various contributions entering the angular momentum and mass-dipole losses.

\begin{table}
	\centering
	\begin{tabular}{| c c c c c c c |}
		\hline 
		 & & Radiative & & Linear static & &  Nonlinear static
		 \\
		\hline
		$J_{ij}$ & $=$ & $\boldsymbol{J}_{ij}$ & $+$  & $\mathcal{J}_{ij}$ & $+$  & $\delta\mathcal{J}_{ij}$
		\\
		& & \checkmark & & \checkmark & &  \checkmark
		\\
		\hline
		$J_{i0}$ & $=$ & $\boldsymbol{J}_{i0}$ & $+$  & $\mathcal{J}_{i0}$ & $+$  & $\delta\mathcal{J}_{i0}$
		\\
		& & \thead{IR divergence} & & \thead{Includes\\Coulombic field} & &  \thead{Collinear\\singularity} \\
		\hline
		$M_{i}$ & $=$ & $\boldsymbol{J}_{i}$ & $+$  & $\mathcal{M}_{i}$ & $+$  & $\delta\mathcal{M}_{i}$
		\\[5pt]
		& & $\boldsymbol{M}_i =\boldsymbol{J}_{i0}-\int t\,\dot P_i dt$ & & $\mathcal{M}_i =\mathcal{J}^{\text{TT}}_{i0}$  & &  $\delta\mathcal{M}_i =\delta \mathcal{J}^{\text{TT}}_{i0}$ \\
		& & \thead{IR finite and\\unambiguous} & & \thead{Excludes\\Coulombic field} & &  \thead{Nonsingular and\\unambiguous}\\
		\hline
	\end{tabular}
	\caption{A short summary of the properties of the various contributions involved in the angular momentum and mass-dipole losses. The $J_{ij}$ components in the center-of-mass frame  are completely regular and unambiguous. Instead, the $J_{i0}$ components inherit an ambiguity under shifts of retarded time due to an IR divergence (tail effect) in $\boldsymbol{J}_{i0}$ and from a collinear singularity in $\delta\mathcal{J}_{i0}$. They can be removed by introducing the subtracted and TT-projected quantity $M_i$. \label{table:properties}}
\end{table}
It will be interesting to extend the analysis presented here, generalizing the discussion in Refs.~\cite{DiVecchia:2022piu,DiVecchia:2023frv} to also include the change in the mechanical angular momentum of the two-body system $\Delta L^{\alpha\beta}=\Delta L_1^{\alpha\beta}+\Delta L_2^{\alpha\beta}$ at $\mathcal{O}(G^4)$. 
This should allow one to check explicitly that the balance law for the total angular momentum $J_{\alpha\beta} = - \Delta L_{\alpha\beta}$ holds, and we will provide a simple argument to this effect in conclusions.
An advantage of having both sides of the balance law under control is that ambiguities associated with the arbitrary time translation mentioned above should cancel out.
It will also be interesting to complete our results by including the full PM $\mathcal{O}(G^4)$ expression for $\boldsymbol{J}_{ij}$ and $\boldsymbol{M}_i$, and by extending them to include angular momentum losses in the presence of tidal and spin effects \cite{Mougiakakos:2022sic,Riva:2022fru,Heissenberg:2022tsn,Heissenberg:2023uvo,DeAngelis:2023lvf,Brandhuber:2023hhl,Aoude:2023dui,Jakobsen:2023hig,Jakobsen:2023pvx,Bohnenblust:2023qmy}.

The paper is organized as follows.
In Section~\ref{sec:eikop}, we briefly review the elastic eikonal exponentiation and its upgrade to the operator version, which draws inspiration from the exponentiation of soft graviton emissions \cite{Weinberg:1964ew,Weinberg:1965nx}, paying particular attention to the role of static and Coulombic modes. In Section~\ref{sec:W} we spell out our conventions for the asymptotic waveform, its TT projection and its multipolar decomposition. In particular, we separate out the contribution due to the static field. Section~\ref{sec:P} is devoted to a review of the energy-momentum loss and serves as an occasion to anticipate several key points that also play  an important role in the following.
Section~\ref{sec:J} contains the new developments involving the angular momentum and is divided into two parts. The first one deals with the calculation of the dynamical emission of angular momentum $\boldsymbol{J}_{ij}$ and mass dipole moment $\boldsymbol{M}_i$, for which we provide explicit results up to $\mathcal{O}(G^4)$ in the small-velocity limit. In the second one, which focuses on static effects, we present the static contributions due to the linear memory effect $\mathcal{J}_{ij}$, $\mathcal{M}_i$, deducing the explicit PM expressions at $\mathcal{O}(G^4)$, and we derive the formula for the nonlinear static contributions $\delta\mathcal{J}_{ij}$, $\delta\mathcal{M}_{i}$ showing in particular that $\delta\mathcal{J}_{ij}$ vanishes at $\mathcal{O}(G^4)$. We end the section by collecting the full expressions for $J_{ij}$ and $M_i$ at $\mathcal{O}(G^4)$ in the PN limit and by verifying the agreement between $J_{ij}$ at this order and the results of Refs.~\cite{Bini:2021gat,Bini:2022enm} up to 2.5PN. The paper also includes two appendices. Appendix~\ref{app:FT} contains useful Fourier transforms from time two frequency domain, while Appendix~\ref{app:upliftTT} provides details about the covariant uplift of the TT formula for the angular momentum loss.

\section{Eikonal operator}
\label{sec:eikop}

The initial state we are interested in is very simple: two  massive particles with masses $m_1$, $m_2$, momenta $p_1$, $p_2$ and total center-of-mass energy\footnote{We work in the mostly plus signature and conventionally take all momenta to be outgoing.} $E=-(p_1^0+p_2^0)$. Since we focus on a scattering process, the final state will contain the same two particles with different momenta ($p_4$ for the particle of mass $m_1$ and $p_3$ for that of mass $m_2$) plus radiation. In the standard perturbative approach, one starts by considering the contribution of the Feynman diagram where a single graviton is exchanged between the two massive particles. 
While we are interested in the four dimensional case, we will need to used dimensional regularization $D=4-2\epsilon$ to make sense of infrared (IR) divergent integrals at intermediate steps.
In momentum space, the contribution of such diagram to the $2\to2$ scattering amplitude reads
\begin{equation}
  \label{eq:1grex}
  {\mathcal A}_0(q^2) = 32 G m_1^2 m_2^2 \left(\sigma^2- \tfrac{1}{D-2}\right) \frac{1}{q^2} + \cdots, 
\end{equation}
where 
\begin{equation}\label{eq:velocityandsigma}
	p_1^\mu = - m_1 v_1^\mu\,,
	\qquad
	p_2^\mu = - m_2 v_2^\mu\,,
	\qquad
    \sigma = -v_1\cdot v_2\ge1
\end{equation}
 and thus $\sigma$ is the relative Lorentz factor between the two particles, while $q=p_1+p_4$ is the momentum carried by the exchanged graviton and the dots in \eqref{eq:1grex} stand for terms that are analytic in $q^2$ in the $q\to 0$ limit. The basic idea of the eikonal approach is that the leading contribution to the full amplitude, where many gravitons are exchanged, is obtained by exponentiating the result in~\eqref{eq:1grex}. This is done by introducing the ``eikonal'' impact parameter\footnote{Another definition of the impact parameter is in terms of the total initial angular momentum $L$: $L=p b$ where $p$ is given in~\eqref{eq:eikft}. The magnitude of these vectors is related by $b=b_e \cos\frac{\Theta}{2}$ with $\Theta$ the deflection angle and also their orientation is different by a relative rotation of $\frac{\Theta}{2}$ (see~\cite{DiVecchia:2023frv} and references therein for a more detailed discussion on this point and on how the transverse Fourier transform~\eqref{eq:eikft} arises). \label{footnote:bebJ}} 
 $b_e$ by means of the Fourier transform
\begin{equation}
  \label{eq:eikft}
  \tilde{\cal A}_0 = \int\! \frac{d^{D-2} q_\perp}{(2\pi)^{D-2}}
  \frac{{\cal A}_0(q^2_\perp)}{4Ep } \, e^{i b_e q_\perp}\,,
\end{equation}
where $E$ is the total center of mass energy as before, $q_\perp$ is the part of $q$ perpendicular to $p_{1,2}$ and $p$ is the spatial momentum in the center-of-mass frame satisfying
\begin{equation}
  \label{eq:Eeq}
  E p =m_1 m_2 \sqrt{\sigma^2-1}\;.
\end{equation}
By using~\eqref{eq:1grex} in~\eqref{eq:eikft} one obtains the leading eikonal
\begin{equation}
  \label{eq:leadeik}
  2\delta_0 \equiv \tilde{\mathcal A}_0 = \frac{2 m_1m_2 G_D \left(\sigma^2- \frac{1}{2\epsilon}\right) \Gamma (-\epsilon) }{\sqrt{\sigma^2-1} (\pi b_e^{-2\epsilon})} \;.
\end{equation}

We can now state the leading eikonal approximation to the S-matrix describing the scattering of two massive particles: it is simply the exponential of~\eqref{eq:leadeik}
\begin{equation}
  \label{eq:SleadEik}
  S(b_e) \simeq e^{2i\delta_0}.
\end{equation}
The Fourier transform back to momentum space $S = \int d^{D-2} b_e e^{-ib_e Q} S(b_e)$ can be well approximated by a saddle point and one obtains a relation between the total momentum exchange $Q$ and the eikonal impact parameter $b_e$, 
\begin{equation}	
Q_\mu = \frac{\partial\,2\delta}{\partial b_e^\mu}\,.
\end{equation}
Notice that $Q$ captures the contribution of a large number of gravitons exchanged between the two massive particles and is a classical quantity, while the momenta of the individual gravitons are of order $\hbar/b_e$. Of course there are classical corrections to the leading eikonal phase discussed so far which promote the phase in~\eqref{eq:SleadEik} to a series of terms $2\delta = 2\delta_0 + 2\delta_1 + \cdots$ where $2\delta_n$  is suppressed by a factor of $(GE/b_e)^n$ with respect to $\delta_0$ and determines as the $(n+1)$PM correction to the classical impulse. The subleading eikonal terms can be extracted from higher-loop corrections to the $2\to2$ amplitude, where more gravitons are exchanged between the two massive particles. In order to do so, one has to subtract from the amplitude the contributions already accounted for by the lower-order terms to isolate the new dynamical information, see~\cite{DiVecchia:2023frv} and references therein for a discussion of this recursive procedure. Here, we just need to mention that $2\delta_1$ is real, so even at 2PM order the elastic eikonal provides a unitary approximation to the $S$-matrix. At 3PM, instead, radiation, i.e.~emission of on-shell gravitons, becomes relevant as highlighted by the fact that $2\delta_2$ develops an imaginary part.

Let us start discussing this breakdown of elastic unitarity by focusing on the emission of soft gravitons, which capture the low frequency limit of the gravitational wave emitted. In this context ``soft'' means that the frequency is small in comparison to the impact parameter: $\omega b_e\ll 1$. A first effect of these soft gravitons is already visible in the 3PM elastic eikonal which develops an \emph{IR-divergent} imaginary part signaling that a purely elastic evolution is impossible as the corresponding $S$-matrix element is suppressed by an infinitely small exponential factor $\exp(-\operatorname{Im}2\delta_2)$. In the soft regime, multi-graviton emissions are governed by the Weinberg soft theorem \cite{Weinberg:1964ew,Weinberg:1965nx,Weinberg:1996kr}. According to this result, the amplitude $\mathcal{A}_{\alpha\to\beta+N}$ for the emission of $N$ gravitons with  soft momenta $k_r$ and polarizations tensors $\varepsilon_{i_r}^{\mu\nu}(k_r)$, with $r=1,\dots,N$, on top of a hard process $\alpha\to\beta$ factorizes as follows,
\begin{equation}\label{eq:WeinbergSoftTheorem}
	\mathcal{A}_{\alpha\to\beta+N} \sim
	 \Bigg[
	 \prod_{r=1}^N \varepsilon^\ast_{i_r\mu\nu}(k_r) F_{\rm tot}^{\mu\nu}(k_r)
	 \Bigg]
	 \mathcal{A}_{\alpha\to\beta}\,,
\end{equation}
as the product of the amplitude $\mathcal{A}_{\alpha\to\beta}$ for the hard process times $N$ copies of the universal factor
\begin{equation}\label{eq:FmunuT}
F_{\rm tot}^{\mu\nu}(k) = \kappa \, \sum_{a} \frac{p_a^\mu p_a^\nu}{p_a\cdot k - i0}\,,
\qquad
\kappa =
\sqrt{8\pi G}
\end{equation}
where $a$ in the last sum runs over all hard states (which can be gravitons themselves). We work with the convention that all momenta $p_a^\mu$ are regarded as formally outgoing, and define the physical (future-directed) momenta $k_a^\mu$ by $p_a^\mu = \eta_a k_a^\mu$ with
\begin{equation}\label{eq:etaadef}
	\eta_a = 
		\begin{cases}
		+1 & \text{if }a\text{ is outgoing}\\
		-1 & \text{if }a\text{ is incoming.}
		\end{cases}
\end{equation}
Soft particles have wavelength much larger than any other length scale involved the process, but still finite, hence $k\neq 0$ and the $-i0$ prescription in \eqref{eq:FmunuT} is irrelevant for them. This is why it was not considered in the original approach of Weinberg. However, extending the formula to``zero-frequency gravitons'' as in \eqref{eq:FmunuT} is useful, as we shall see, in order to capture static contributions to gravitational observables such as the angular momentum.

The $N$-graviton soft theorem \eqref{eq:WeinbergSoftTheorem} essentially states that, to leading order in the soft limit, graviton emissions are completely uncorrelated and produce a coherent superposition. A compact way to encode this is to introduce the operator
\begin{equation}\label{eq:opF1}
	\hat{S}_{s.r.} =
	e^{2i\tilde{\delta}(b_e)}
	\exp\left[
	\int_k \theta(\omega^\ast-\omega)
	\left(
	F(k) a^\dagger(k)
	-
	F^{\ast}(k) a(k)
	\right)
	\right],
\end{equation}
where $2\tilde{\delta}$ is obtained from the real part of the elastic eikonal discussed above,\footnote{
	In order to define $2\tilde\delta$, it is possible to link \eqref{eq:opF1} to in and out states dressed by soft gravitons.
	See \cite{DiVecchia:2022piu,DiVecchia:2023frv} for the precise mechanism through which 3PM Radiation-Reaction enters via the saddle-point conditions.} $a_{\mu\nu}(k)$, $a_{\mu\nu}^\dagger(k)$ are graviton creation and annihilation operators in de Donder gauge, which obey
\begin{equation}
	2\pi\theta(k^0)\delta(k^2)
	[
	a_{\mu\nu}(k),  a_{\rho\sigma}^\dagger(k')
	]
	=
	(2\pi)^D\delta^{(D)}(k-k')\,
	\tfrac{1}{2}
	\left(
	\eta_{\mu\rho} \eta_{\nu\sigma}
	+
	\eta_{\mu\sigma} \eta_{\nu\rho}
	-
	\tfrac{2}{D-2}
	\,
	\eta_{\mu\nu} \eta_{\rho\sigma}
	\right).
\end{equation}
We also employed the shorthand notation
\begin{equation}
	\int_k = \int \frac{d^4k}{(2\pi)^4}\,2\pi \theta(k^0) \delta(k^2)\,,
	\qquad
	F a^\dagger = F_{\mu\nu} 
	\left(
	\eta_{\mu\rho} \eta_{\nu\sigma}
	-
	\tfrac{1}{2}
	\,
	\eta_{\mu\nu} \eta_{\rho\sigma}
	\right)
	a^\dagger_{\rho\sigma} \;,
\end{equation}
where, as a first step, we make a further simplifying assumption: we consider only the soft gravitons emitted by the scalar external states representing the black holes, so that $F^{\mu\nu}$ has the same form as in~\eqref{eq:FmunuT}, but with the sum restricted to the {\em massive} states labeled by $a_m$
\begin{equation}\label{eq:Fmunu}
F^{\mu\nu}(k) = \kappa \, \sum_{a_m} \frac{p_{a_m}^\mu p_{a_m}^\nu}{p_{a_m}\cdot k - i0}\,.
\end{equation}
In this way, $\hat S_{s.r.}|0\rangle$ ``dresses'' the vacuum with a coherent superposition of gravitons with frequency $\omega<\omega^\ast$ emitted by the massive states. We will come back to the full soft expression in~\eqref{eq:FmunuT} when discussing the role of the nonlinear memory effect, see Eq.~\eqref{eq:Ftot} and below. In the operator formalism just introduced the elastic process is described by $\langle 0| \hat{S}_{s.r.}|0 \rangle$ and the divergent imaginary part of the $2\delta_2$ mentioned above is obtained by normal ordering the creation and annihilation operators in~\eqref{eq:opF1}~\cite{DiVecchia:2021ndb}.

They key physical idea that graviton emissions should generically be coherent in the classical limit, at least order by order in the small-deflection limit, has motivated the proposal of a more general eikonal operator~\cite{Cristofoli:2021jas,DiVecchia:2022piu}, 
\begin{equation}\label{eq:Shat}
	\hat{S} =
	e^{2i\tilde\delta(b_e)}
	\exp\left[
	i
	\int_k
	\left(
	\tilde{\tau}^{\mu\nu}(k) a_{\mu\nu}^\dagger(k)
	+
	\tilde{\tau}_{\mu\nu}^{\ast}(k) a^{\mu\nu}(k)
	\right)
	\right],
\end{equation}
which should describe the final state of a classical collision even beyond the soft limit. 
While $2\tilde\delta(b_e)$ is determined by the elastic amplitude, the new ingredient  $\tilde{\tau}_{\mu\nu}$ is the frequency-domain field, which receives two types of contributions: a soft/static one and a hard/dynamical one. Letting $\omega^\ast$ denote the cutoff frequency marking the separation between the two,
\begin{equation}\label{eq:splitting}
	\tilde{\tau}^{\mu\nu}(k)
	= \lim_{\omega^\ast \to 0} \left[
	-i
	\theta(\omega^\ast-\omega)
	F^{\mu\nu}(k)
	+
	\theta(\omega-\omega^\ast)
	\tilde W^{\mu\nu}(k) \right].
\end{equation}
Here $F^{\mu\nu}(k)$ is the static contribution, which is captured by \eqref{eq:Fmunu},
while $\tilde{W}^{\mu\nu}$ 
is the dynamical contribution. Eventually, we let $\omega^\ast\to 0$, so the results captured by the first term in~\eqref{eq:splitting} are due to the terms in the asymptotic metric fluctuation that are independent of this splitting (and in the time domain, independent of $t$).

The dynamical contribution $\tilde{W}^{\mu\nu}$ can be written as the Fourier transform of a momentum space ``kernel'' $W^{\mu\nu}$,
\begin{align}
	\nonumber
	\tilde{W}^{\mu\nu}(b_1,b_2,k) &= \int\frac{d^Dq_1}{(2\pi)^D}\frac{d^Dq_2}{(2\pi)^D} (2\pi)^D \delta^{(D)}(q_1+q_2+k) 2\pi\delta(2p_1\cdot q_1)2\pi\delta(2p_2\cdot q_2)\\
	&\times e^{ib_1\cdot q_1+i b_2\cdot q_2} W^{\mu\nu}(q_1,q_2,k)\,,
	\label{eq:FT5}
\end{align}
which can be derived from the classical limit of the $5$-point amplitude ${\mathcal A}^{(5)}$ describing the scattering of the massive state with the emission of an extra on-shell graviton. The tree-level approximation $\mathcal{A}_0^{(5)}$ for this amplitude was discussed in~\cite{Goldberger:2016iau,Luna:2017dtq,Mogull:2020sak}, and then $W_0^{\mu\nu} = \mathcal{A}_0^{(5)\mu\nu}$.
The one-loop expression for the amplitude, $\mathcal{A}_1^{(5)\mu\nu}$, was obtained in~\cite{Brandhuber:2023hhy,Herderschee:2023fxh,Georgoudis:2023lgf}. Exactly as in the elastic case, the NLO correction to the waveform kernel $W_1^{\mu\nu}$ is obtained from the amplitude $\mathcal{A}_1^{(5)\mu\nu}$ after an appropriate subtraction~\cite{Caron-Huot:2023vxl,Georgoudis:2023eke}. Because of its origin, the NLO contribution $W_1^{\mu\nu}$ thus contains both real and imaginary parts, contrary to the leading term which is purely real. This is the origin of some new subleading contributions in the calculation of the angular momentum loss as discussed in Sect.~\ref{sec:P}.
Being derived from gauge-invariant amplitude,  the dynamical contribution  obeys the transversality condition
\begin{equation}\label{eq:upliftW}
	k_\mu
	\tilde{W}^{\mu\nu}(k)
	=0\,.
\end{equation}

\section{Waveform}
\label{sec:W}

The eikonal operator in Eq.~\eqref{eq:Shat}
determines the gravitational field sourced by the collision via \cite{Kosower:2018adc,Cristofoli:2021vyo}
\begin{equation}
	h_{\mu\nu}(x) = g_{\mu\nu}(x)-\eta_{\mu\nu} = \frac{1}{2\kappa}\, \langle\text{in}| \hat{S}^\dagger \int_k \left(e^{ik\cdot x} a_{\mu\nu}(k) +  e^{-ik\cdot x} a^\dagger_{\mu\nu}(k) \right) \hat{S}  |\text{in}\rangle\,.
\end{equation}
We then consider the limit of large radial distance $r\to\infty$ for fixed retarded time $t$ and angular direction $n^\mu = (1,\hat n)$, by letting $x^\mu = (t+r,r \hat{n})$.
Focusing on the leading asymptotic limit of the metric fluctuation, we then obtain \cite{Cristofoli:2021vyo,DiVecchia:2023frv}
\begin{equation}\label{eq:hmunu}
	h_{\mu\nu}(x) = g_{\mu\nu}(x)-\eta_{\mu\nu} = \frac{\tau_{\mu\nu}(t,n)}{r} + \cdots\;,
\end{equation}
where we can move from the frequency-domain to the time-domain representation using
\begin{equation}\label{eq:waveform}
	\tau_{\mu\nu}(t,n)=\frac{4G}{\kappa} \left[\int_{0}^{+\infty}
          e^{-i\omega t}\, \tilde{\tau}_{\mu\nu}(\omega n)\,\frac{d\omega}{2\pi} +
          \int_{0}^{+\infty} e^{i\omega t}\, \tilde{\tau}^\ast_{\mu\nu}(\omega n)\,
          \frac{d\omega}{2\pi}\right].
\end{equation}

For practical applications, one can also focus on the gauge-invariant content of the dynamical waveform $\tilde{W}^{\mu\nu}$ by taking its transverse-traceless (TT) projection,
\begin{equation}\label{eq:TTprojection}
	\tilde{w}_{\mu\nu}(k) 
	= \Pi_{\mu\nu,\rho\sigma}(k) 
	\tilde{W}^{\rho\sigma}(k)\,,
\end{equation}
where the TT projector $\Pi_{\mu\nu,\rho\sigma}(k)$ can be defined as follows,
\begin{align}
	\label{eq:TTprojector}
	 \Pi_{\mu\nu,\rho\sigma}
	&= \frac{1}{2}\left(
	 \Pi_{\mu\rho} \Pi_{\nu\sigma}
	 +
	 \Pi_{\mu\sigma} \Pi_{\nu\rho}
	 -
	 \frac{2}{D-2}
	 \Pi_{\mu\nu} \Pi_{\rho\sigma}
	 \right),
	 \\
	 \label{eq:Tprojector}
	 \Pi_{\mu\nu} &= \eta_{\mu\nu} + \lambda_{\mu} k_{\nu} + \lambda_{\nu} k_{\mu}
\end{align}
in terms of the arbitrary reference vector $\lambda^\mu$ obeying $\lambda^2=0$, $\lambda\cdot k=-1$.
One often considers expressions defined in the center-of-mass frame, 
where 
\begin{equation}
	\label{eq:ndecomp}
	k^\mu = \omega (1, \hat n)\,,
	\qquad
	\hat{n} = (n_x, n_y, n_z) = (\sin\theta\cos\phi,\sin\theta\sin\phi,\cos\theta)
\end{equation}
and then it is natural to choose
\begin{equation}
	\label{eq:lambdadecomp}
	\lambda^\mu = \frac{1}{2\omega} (1, -\hat n)\,,
\end{equation}
so that the transverse projector is purely spatial,
\begin{equation}
	\label{eq:Pimunusimplified}
	\Pi_{\mu\nu} = \left(\begin{matrix}
0 & 0 \\
0 & \delta_{ij} - n_i n_j
	\end{matrix}\right).
\end{equation}
Since the dynamical TT waveform reduces in this way to a traceless spatial tensor tangent to the two-sphere, it can be equivalently presented as a multipolar decomposition \cite{Thorne:1980ru,Blanchet:1985sp,Blanchet:2013haa}, 
\begin{equation}\label{eq:multipoleDecFreq}
	\frac{1}{\kappa}\,\tilde{w}_{ab}
	=
	\sum_{\ell=2}^\infty\frac{1}{\ell!}
	\Bigg[
	n_{L-2}\, \boldsymbol{\mathrm{U}}_{ijL-2}(u)
	-\frac{2\ell}{\ell+1}\,n_{cL-2}\epsilon_{cd(i} \, \boldsymbol{\mathrm{V}}_{j)dL-2}(u)
	\Bigg] \Pi_{ij,ab}\,,
\end{equation}
where $\boldsymbol{\mathrm{U}}_{L}$ and $\boldsymbol{\mathrm{V}}_L$ are symmetric trace-free (STF) tensors, we include the factor of $\tfrac12$ in the symmetrization $v_{(i} w_{j)} = \frac{1}{2} (v_i w_j + v_j w_i)$, and $u$ denotes the dimensionless frequency variable 
\begin{equation}
	u = \frac{\omega b}{p_\infty}\,.
\end{equation}
The multipole decomposition \eqref{eq:multipoleDecFreq} is particularly useful in the PN limit, since order by order in the small-velocity expansion the sum over $\ell$ truncates to the first few multipoles. Notice that the multipoles introduce in~\eqref{eq:multipoleDecFreq} differ in one aspect from those that are more commonly employed in the PN literature~\cite{Blanchet:2013haa} as they do not contain the static $G$-independent part. As mentioned above, those contributions are encoded in $F^{\mu\nu}$, see~\eqref{eq:splitting}.

We choose to align our coordinate axes in the center-of-mass frame according to the following conventions, 
\begin{equation}\label{eq:refvec}
	b_e^{\mu} =b_1^\mu-b_2^\mu = b_e (0,1,0,0)\,,\qquad
	e^{\mu} = (0,0,1,0)\,,
\end{equation}
with $e^\mu$ oriented along the spatial projection of the ``average'' velocities $\tilde u_{1,2}^\mu$, satisfying $\tilde{u}_1^2=-1$, $\tilde{u}_2^2=-1$, defined by the $\mathcal{O}(G^2)$-accurate dynamics
\begin{equation}\label{eq:G2accurate}
	\begin{split}
		p_1^\mu &= - m_1 \tilde u_1^\mu + \frac{1}{2}\,Q^\mu\,,\qquad
		p_4^\mu =  m_1 \tilde u_1^\mu + \frac{1}{2}\,Q^\mu\,,\\
		p_2^\mu &= -  m_2 \tilde u_2^\mu - \frac{1}{2}\,Q^\mu\,,\qquad
		p_3^\mu =  m_2 \tilde u_2^\mu - \frac{1}{2}\,Q^\mu\,.
	\end{split}
\end{equation}
We  fix the boost-freedom in the choice of reference frame by $m_1 \tilde {u}_1^{i} + m_2 \tilde{u}_2^{i} = 0$, which
	at $\mathcal{O}(G^2)$  is equivalent to $p_1^{i} + p_2^{i} = 0$. As discussed  below Eq.~\eqref{eq:Q1Q2P}, we will keep the same choice also at the next order in $G$.
Furthermore, we fix the origin and thus the translation-freedom by 
\begin{equation}\label{eq:CoMdefO}
E_1 b_1^\mu + E_2 b_2^\mu = 0\,,
\end{equation}
where $E_1$, $E_2$ are the energies of the particles in the center-of-mass frame. Again, at this order, Eq.~\eqref{eq:CoMdefO} can be stated equivalently in terms of incoming or ``average'' impact parameters, since the two only differ by a rotation.
We also introduce the ``velocity'' $p_\infty$, the total mass $m$ and the symmetric mass-ratio $\nu$ according to
\begin{equation}
	\sigma = \sqrt{1+p^2_\infty}\,,
	\qquad
	m = m_1 + m_2\,,
	\qquad
	\nu = \frac{m_1 m_2}{m^2}\,,
\end{equation}
so that the PN limit corresponds to taking $p_\infty\to0$.

\section{Energy-momentum loss}
\label{sec:P}

The energy-momentum loss is insensitive to the static effects and thus equals the one radiated via emission of gravitational waves $P^\alpha = \boldsymbol{P}^\alpha$, taking the form
\cite{Herrmann:2021lqe}
\begin{equation}
	\label{eq:PmuGaugeInv}
	{P}^\alpha =
	\int_k k^\alpha \rho(k)\,,
\end{equation}
where we introduced the following notation
\begin{equation}\label{eq:spectralemissionrate}
	\rho 
	= 
	 \tilde{W}^{\ast}_{\mu\nu} 
	\left(
	\eta^{\mu\rho}
	\eta^{\nu\sigma}
	-
	\tfrac{1}{D-2}\,
	\eta^{\mu\nu}
	\eta^{\rho\sigma}
	\right)
	\tilde{W}_{\rho\sigma}
\end{equation}
for the spectral emission rate, i.e.~the phase space density of graviton emissions.
Note that \eqref{eq:PmuGaugeInv} and \eqref{eq:spectralemissionrate} are invariant under translations and thus, for the scattering of scalar objects which we focus on here, $\rho$ is a function of the following invariants,
\begin{equation}
	\rho = \rho(b_e^2,b_e\cdot k, \tilde{u}_1\cdot k,\tilde{u}_2\cdot k)\,.
\end{equation}
Let us also note that $\rho$ starts at order $\mathcal{O}(G^3)$ in the PM expansion, when $W^{\mu\nu}$ is approximated with the tree-level amplitude $\mathcal{A}_0^{(5)\mu\nu}$,
\begin{equation}\label{eq:rho0}
	\rho = \tilde{\mathcal{A}}_0^{(5)\ast\mu\nu} \left(
	\eta_{\mu\rho}\eta_{\nu\sigma}
	-\tfrac{1}{D-2}\,
	\eta_{\mu\nu}
	\eta_{\rho\sigma}
	\right)
	\tilde{\mathcal{A}}_0^{(5)\rho\sigma} + \mathcal{O}(G^4) = \rho_0 + \mathcal{O}(G^4) \,.
\end{equation}
An important property of the leading-order spectral rate $\rho_0$ is that, owing to the reality of the tree-level amplitude in momentum space and to the Fourier transform~\eqref{eq:FT5}, 
\begin{equation}\label{eq:rho0isEVEN}
	\tilde{\mathcal{A}}_0^{(5)\mu\nu}(-b_1,-b_2,k) =\tilde{\mathcal{A}}_0^{(5)\ast\mu\nu} (b_1,b_2,k)\,,
	\qquad
	\rho_0 = \rho_0\big|_{b_e\cdot k\to-b_e\cdot k} 
\end{equation}
that is, $\rho_0$  is an \emph{even} function of $b_e \cdot k$.

Thanks to the transversality condition \eqref{eq:upliftW}, one immediately sees that Eq.~\eqref{eq:PmuGaugeInv} is equivalent to the following expression involving the TT waveform $\tilde{w}^{ab}$,
\begin{equation}
	\label{eq:PmuTT}
	P^\alpha 
	=
	\int_k k^\alpha\,\tilde{w}^{\ast}_{ab} 
	\tilde{w}_{ab}\,.
\end{equation}
Following the decomposition \eqref{eq:ndecomp} in the center-of-mass frame we thus have
\begin{align}
	\label{eq:P0oneline}
	{\kappa^2} P^0 &\equiv \kappa^2 E_\text{rad} = G \int_0^\infty \frac{d\omega}{\pi} \oint \frac{d\Omega}{2\pi}\,\omega^2\tilde{w}^{\ast}_{ab} 
	\tilde{w}_{ab}\,,\\
	\label{eq:Pioneline}
	{\kappa^2} P^i &= G \int_0^\infty \frac{d\omega}{\pi} \oint \frac{d\Omega}{2\pi}\,\omega^2 n^i \tilde{w}^{\ast}_{ab} 
	\tilde{w}_{ab}\,,
\end{align}
which can be also obtained by integrating the following emission spectra expressed in terms of multipole moments,
\begin{equation}
	\label{eq:dEdo}
	\frac{dE_\text{rad}}{d\omega} = \frac{G}{\pi} \sum_{\ell=2}^{+\infty} \left[\frac{(\ell+1)(\ell+2)\,\omega^2  \boldsymbol{\mathrm{U}}^*_L \boldsymbol{\mathrm{U}}_L}{(\ell-1) \ell \ell !(2 \ell+1) ! !}  + \frac{4 \ell(\ell+2)\,\omega^2  \boldsymbol{\mathrm{V}}^*_L \boldsymbol{\mathrm{V}}_L}{(\ell-1)(\ell+1) !(2 \ell+1) ! !}\right],
\end{equation}
and
\begin{equation}
	\label{eq:dPdo}
	\begin{split}
	\frac{d P_i}{d\omega} 
	= 
	\frac{G}{\pi} \sum_{\ell=2}^{\infty}\operatorname{Re}
	&\Bigg[ \frac{2(\ell+2)(\ell+3)\, \omega^2}{\ell(\ell+1) !(2 \ell+3) ! !} \boldsymbol{\mathrm{U}}_{i L}^* \boldsymbol{\mathrm{U}}_L +\frac{8(\ell+3)\, \omega^2 \boldsymbol{\mathrm{V}}^*_{i L} \boldsymbol{\mathrm{V}}_L }{(\ell+1) !(2 \ell+3) ! !} 
	\\
	&+ \frac{8(\ell+2) \, \omega^2 \,\epsilon_{i a b} \boldsymbol{\mathrm{U}}^*_{a L-1} \boldsymbol{\mathrm{V}}_{b L-1}}{(\ell-1)(\ell+1) !(2 \ell+1) ! !}  \Bigg],
	\end{split}
\end{equation}
or the time-domain fluxes (see the useful relation \eqref{eq:dotf_dotg})
\begin{equation}\label{eq:time-domain-flux-P}
	\frac{dE_\text{rad}}{dt} = \frac{1}{32\pi G}  \oint {d\Omega}\, \dot{w}^{ab} 
	\dot{w}^{ab}\,,\qquad
	\frac{dP^i}{dt} = \frac{1}{32\pi G}  \oint {d\Omega}\, \dot{w}^{ab} 
	\dot{w}^{ab} n^i\,.
\end{equation}

Substituting into \eqref{eq:dEdo}, \eqref{eq:dPdo} the post-Newtonian multipoles obtained in \cite{Georgoudis:2024pdz,Bini:2024rsy} and integrating over the frequency, one finds 
\begin{equation}\label{eq:Erad}
	\begin{split}
	E_\text{rad}
	&=\frac{G^3\pi m^4}{b^3}\, p_\infty\nu^2
	\left[
	\frac{37}{15}
	+
	\left(\frac{1357}{840}-\frac{37 \nu }{30}\right) p_\infty^2
	+\mathcal O(p_\infty^4)
	\right]
	\\
	&+
	\frac{G^4m^5}{b^4p_\infty}
	\nu^2
	\left[
	\frac{1568}{45}
	+
	\left(\frac{18608}{525}-\frac{1136}{45}\nu\right) p_\infty^2
	+\mathcal O(p_\infty^4)
	\right]
	\\
	&+
	\frac{G^4m^5}{b^4}p_\infty^2
	\nu^2
	\left[
	\frac{3136}{45}
	+
	\left(\frac{1216}{105}-\frac{2272}{45}\nu\right) p_\infty^2
	+\mathcal O(p_\infty^4)
	\right]
	\\
	&+\mathcal O(G^5)\,.
	\end{split}
\end{equation}
The expression for the first line with exact dependence on $\sigma=\sqrt{1+p_\infty^2}$, the 3PM contribution, is given by Eq.~(10) of Ref.~\cite{Herrmann:2021lqe}. Coming to the nonvanishing components of $P_i$, recalling the choice of axes \eqref{eq:refvec} we have
\begin{equation}\label{eq:Prad1}
	\begin{split}
		P_x
		&=
		\frac{G^4m^5}{b^4}\,p_\infty^3
		\nu^2 \sqrt{1-4\nu}
		\left(\frac{1491}{400} - \frac{26757}{5600}p_\infty^2+\mathcal O(p_\infty^4)\right) \pi
		\\
		&+\mathcal O(G^5)\,.
	\end{split}
\end{equation}
and
\begin{equation}\label{eq:Prad2}
	\begin{split}
	P_y
	&=\frac{G^3\pi m^5}{b^3} \,
	p_\infty^2
	\nu^2 \sqrt{1-4\nu}
	\left[
	-\frac{37}{30}
	+
	\left(
	\frac{37}{60}\nu
	-
	\frac{839}{1680}
	\right)
	p_\infty^2
	+\mathcal O(p_\infty^4)
	\right]
	\\
	&+
	\frac{G^4m^5}{b^4}
	\nu^2 \sqrt{1-4\nu}
	\left[
	-\frac{64}{3}
	+
	\left(
	\frac{32}{3}\nu
	-
	\frac{1664}{175}
	\right)
	p_\infty^2
	+\mathcal O(p_\infty^4)
	\right]
	\\
	&+
	\frac{G^4m^5}{b^4}\,p_\infty^3
	\nu^2 \sqrt{1-4\nu}
	\left[
	-\frac{128}{3} +
	\left(
	\frac{64}{3}\nu- \frac{192}{75}
	\right)p_\infty^2
	+\mathcal O(p_\infty^4)
	\right]
	\\
	&+\mathcal O(G^5)\,.
	\end{split}
\end{equation}
Eqs.~\eqref{eq:Erad}, \eqref{eq:Prad1}, \eqref{eq:Prad2} are in precise agreement with the results obtained by Refs.~\cite{Herrmann:2021lqe,Herrmann:2021tct,Bini:2021gat,Dlapa:2022lmu,Bini:2022enm}. 
Note also that the radiated momentum vanishes, ${P}_i=0$, for the equal-mass case, $m_1=m_2$, as can be expected by symmetry considerations. 

The $\mathcal{O}(G^3)$ result for $P^\alpha$ was obtained for generic velocities in \cite{Herrmann:2021lqe,Herrmann:2021tct}  and let us emphasize that the vanishing of the component of the radiated momentum along the impact parameter, $P_x$, at that order follows from the fact that $\rho_0$ is even in $b_e\cdot k$, \eqref{eq:rho0isEVEN}, and thus
\begin{equation}
	\int_k b_e\cdot k \,\rho_0 = 0\,.
\end{equation}
At the next order, $\mathcal{O}(G^4)$, the property \eqref{eq:rho0isEVEN} no longer holds and indeed the $P_x$ component is no longer zero \eqref{eq:Prad1} as it involves the imaginary part of the one-loop waveform kernel. 
In particular, \eqref{eq:Prad1} is sensitive to the (unique) subtraction of $\epsilon/\epsilon$ terms induced by the resummation of infrared divergences discussed in \cite{Georgoudis:2024pdz,Bini:2024rsy}.

The vanishing of $P_z$ is also obvious in the scalar case. This follows from the fact that $\rho$ is independent of $\zeta\cdot k$, where $\zeta^\mu$ is the unit vector aligned with the $z$ axis, which is orthogonal to the scattering plane in our conventions \eqref{eq:refvec}. When including spin effects \cite{Riva:2022fru}, a nonzero $P_z$ component can arise in the misaligned case.

\section{Angular momentum loss}
\label{sec:J}

The angular momentum/mass dipole that the two-body system dissipates can be computed from the manifestly Poincar\'e covariant formula
\cite{Manohar:2022dea,DiVecchia:2022owy}
\begin{equation}\label{eq:JmunuGaugeInv}
	i J_{\alpha\beta} =
	\int_k
	\left[
	\left(
	\eta^{\mu\rho}
	\eta^{\nu\sigma}
	-
	\tfrac{1}{D-2}\,
	\eta^{\mu\nu}
	\eta^{\rho\sigma}
	\right)
	\tilde{\tau}_{\mu\nu}^{\ast}
	k_{[\alpha}^{\phantom{\mu}}
	\frac{\overset{\leftrightarrow}{\partial} \tilde{\tau}_{\rho\sigma}}{\partial k^{\beta]}}
	+2
	\eta^{\mu\nu}
	\tilde{\tau}_{\mu[\alpha}^{\ast}
	\tilde{\tau}_{\beta]\nu}^{\phantom{\ast}}
	\right],
\end{equation}
where the antisymmetrization is defined by $v_{[\alpha} w_{\beta]}=v_{\alpha} w_{\beta}-v_{\beta} w_{\alpha}$ and
\begin{equation}\label{eq:leftrightarrow}
	f \overset{\leftrightarrow}{\partial} g = \frac{1}{2}
	\left(
	f \, {\partial} g
	-
	g \,  {\partial} f
	\right).
\end{equation}
Eq.~\eqref{eq:JmunuGaugeInv} receives two types of contributions \eqref{eq:splitting}, which we denote as follows,
\begin{equation}
	J_{\mu\nu}
	=
	\boldsymbol{J}_{\mu\nu}
	+
	\mathcal{J}^{\rm tot}_{\mu\nu}
	\,.
\end{equation}
The first one, $\boldsymbol{J}_{\mu\nu}$, is the dynamical or radiative contribution, which is the one  carried away by the gravitational field, while the second one, $\mathcal{J}^{\rm tot}_{\mu\nu}$,  is due to the interaction with the static field.

\subsection{Radiative contribution}
\label{ssec:boldJ}

We begin from the radiative contribution to \eqref{eq:JmunuGaugeInv}, 
\begin{equation}\label{eq:JmunuGaugeInvw}
	i \boldsymbol{J}_{\alpha\beta} =
	\int_k
	\left[
	\left(
	\eta^{\mu\rho}
	\eta^{\nu\sigma}
	-
	\tfrac{1}{D-2}\,
	\eta^{\mu\nu}
	\eta^{\rho\sigma}
	\right)
	\tilde{W}_{\mu\nu}^{\ast}
	k_{[\alpha}^{\phantom{\mu}}
	\frac{\overset{\leftrightarrow}{\partial} \tilde{W}_{\rho\sigma}}{\partial k^{\beta]}}
	+2
	\eta^{\mu\nu}
	\tilde{W}_{\mu[\alpha}^{\ast}
	\tilde{W}_{\beta]\nu}^{\phantom{\ast}}
	\right].
\end{equation}
Thanks to the exact transversality property \eqref{eq:upliftW}, Eq.~\eqref{eq:JmunuGaugeInvw} can be expressed as the following integral quadratic in the dynamical component of the TT waveform \cite{DiVecchia:2022owy},
\begin{equation}\label{eq:JmunuTT}
i \boldsymbol{J}_{\alpha\beta} =
\int_k
\left[
\tfrac{1}{2}
\left(
\tilde{w}_{\mu\nu}^{\ast}
k_{[\alpha}^{\phantom{\mu}}
\frac{\partial \tilde{w}^{\mu\nu}}{\partial k^{\beta]}}
-
\tilde{w}^{\mu\nu}
k_{[\alpha}^{\phantom{\mu}}
\frac{\partial \tilde{w}^\ast_{\mu\nu}}{\partial k^{\beta]}}
\right)
+2
\tilde{w}_{\mu[\alpha}^{\ast}
\tilde{w}_{\beta]}^{\mu}
\right].
\end{equation}
In particular, the equivalence between \eqref{eq:JmunuTT} and \eqref{eq:JmunuGaugeInvw} holds thanks to the fact that the reference vector $\lambda^\mu$ appearing in the TT projector \eqref{eq:TTprojector} actually drops out from \eqref{eq:JmunuTT}.
In the following, we use the ``TT formula'' \eqref{eq:JmunuTT} and the explicit expressions for the waveforms $\tilde{w}_{\mu\nu}$ to calculate $\boldsymbol{J}_{\alpha\beta}$ in the center-of-mass frame following the choice of coordinate axes in \eqref{eq:refvec}.

We consider first the purely spatial components $\boldsymbol{J}_{ij}$, which represent the angular momentum proper in the center-of-mass frame.
Having chosen the reference vector in the TT projector as in \eqref{eq:lambdadecomp} so that $\tilde{w}_{0 \mu}=0$ by \eqref{eq:Pimunusimplified}, we have
\begin{equation}\label{eq:JijTT}
	i \boldsymbol{J}_{ij} =
	\int_k
	\left[
	\tfrac{1}{2}
	\left(
	\tilde{w}_{ab}^{\ast}
	k_{[i}^{\phantom{i}}
	\frac{\partial \tilde{w}_{ab}}{\partial k^{j]}}
	-
	\tilde{w}_{ab}
	k_{[i}^{\phantom{j}}
	\frac{\partial \tilde{w}^\ast_{ab}}{\partial k^{j]}}
	\right)
	+2
	\tilde{w}_{a[i}^{\ast}
	\tilde{w}_{j]a}
	\right],
\end{equation}
which is the standard formula used in the general relativity literature \cite{Thorne:1980ru,Maggiore:2007ulw,Damour:2020tta}.
Introducing spherical variables in the integration,
\begin{equation}\label{eq:changeofv}
	k^0 = \omega = |\vec{k}\,|\,,
	\qquad
	k^i = \omega n^i 
\end{equation}
with $n^i$ the standard unit vector, as in \eqref{eq:ndecomp},
\begin{equation}
	\hat n =(\sin\theta \cos\phi,\sin\theta \sin\phi,\cos\theta)\,,
	\qquad
	x^A = (\theta,\phi)\,,
\end{equation}
it is convenient to note that the Jacobian matrix for the spatial part reads
\begin{equation}
	{M}\indices{^i_j} = \left( n^i, \omega \partial_A n^i \right)\,,\qquad \left({M}^{-1}\right)\indices{^j_i} = 
	\left(
	\begin{matrix}
		n_i\\
		\omega^{-1} \gamma^{AB}\partial_B n_i
	\end{matrix}\right)
\end{equation}
where $\gamma_{AB} = (\partial_A \hat n)\cdot(\partial_B \hat n)=\text{diag}(1,(\sin\theta)^2)$ is the metric on the unit sphere and $\gamma^{AB}$ denotes its inverse,
and thus
\begin{equation}
\omega\,	\frac{\partial}{\partial k^ i} = n_i \,\omega \partial_\omega  + \gamma^{AB} \partial_B n_i \partial_A\,.
\end{equation}
Using Eqs.~\eqref{eq:changeofv} and following, we can recast \eqref{eq:JijTT} as
\begin{equation}\label{eq:Jij_twolines}
	\begin{split}
		\kappa^2 \boldsymbol{J}^{ij} 
		&= 
		G \int_0^\infty \frac{d\omega}{i\pi}\oint \frac{d\Omega}{2\pi}
		\tilde{w}^\ast_{ab} \partial_A \tilde{w}_{ab}
		\,
		\omega\,
		\gamma^{AB} n^{[i}\partial_B n^{j]}
		\\
		&+
		2G \int_0^\infty \frac{d\omega}{i\pi}\oint \frac{d\Omega}{2\pi}
		\tilde{w}^{\ast a[i} \tilde{w}^{j]a}\,\omega\,,
	\end{split}
\end{equation}
where we have used $D_A D_B n^i = -\gamma_{AB} n^i$ to integrate by parts with respect to $\partial_B$.
Explicitly,
\begin{subequations}
	\begin{align}
		\gamma^{AB} n^{[x}\partial_B n^{y]} \partial_A &= \partial_\phi \,,\\
		\gamma^{AB} n^{[y}\partial_B n^{z]} \partial_A &= - \sin\phi\,\partial_\theta-\frac{\cos\theta}{\sin\theta}\,\cos\phi\,\partial_\phi \,,\\ 
		\gamma^{AB} n^{[z}\partial_B n^{x]} \partial_A &= +\cos\phi\,\partial_\theta-\frac{\cos\theta}{\sin\theta}\,\sin\phi\,\partial_\phi \,.
	\end{align}
\end{subequations}
For completeness, let us mention that, using  \eqref{eq:dotf_g} and taking into account the factors in \eqref{eq:waveform}, one can rewrite \eqref{eq:Jij_twolines} as the integral over time of the following time-domain flux \cite{Damour:2020tta},\footnote{The differential operator appearing in \eqref{eq:Jijdot} can be also rewritten as $n^{[i}\partial_B n^{j]} \gamma^{AB} \partial_A = x^{[i}\partial^{j]}$ in terms of embedding space coordinates $x^i= r \,n^i$.}
\begin{equation}\label{eq:Jijdot}
		\frac{d \boldsymbol{J}_{ij} }{dt}
		= 
		-
		\frac{1}{32\pi G} \oint d\Omega
		\left(
		\dot w^{ab}  \partial_A w^{ab} \gamma^{AB} n^{[i} \partial_B
		n^{j]}
		+
		2 \dot w^{a[i} w^{j]a}\right).
\end{equation}
Similarly, one finds the following expression fore the emission spectrum in frequency-domain, written in terms of multipoles,
\begin{equation}\label{eq:JijMultipoles}
	\frac{d\boldsymbol{J}_{ij}}{d\omega}
	=
	\frac{G\omega}{\pi} \operatorname{Im} \sum_\ell \left[
	\frac{(\ell+1)(\ell+2)\boldsymbol{\mathrm{U}}^{\phantom{\ast}}_{L[i}\boldsymbol{\mathrm{U}}^\ast_{j]L}}{(\ell-1) \ell! (2\ell+1)!!}
	+
	\frac{4\ell^2(\ell+2) \boldsymbol{\mathrm{V}}^{\phantom{\ast}}_{L[i}\boldsymbol{\mathrm{V}}^\ast_{j]L}}{(\ell-1)(\ell+1)! (2\ell+1)!!}
	\right].
\end{equation}
The $xy$ component also affords a particularly simple representation
\begin{equation}
\kappa^2 \boldsymbol{J}_{xy} 
= 
G \int_0^\infty \frac{d\omega}{i\pi}\oint \frac{d\Omega}{2\pi}
\sum_{s=+,\times}
\tilde{w}_s^{\ast} \partial_\phi \tilde{w}^{\phantom{\ast}}_{s}
\,
\omega\,
\end{equation} 
in  terms of the decomposition given by
\begin{equation}
	\tilde{w}^{ab} = \sum_{s=+,\times} \tilde{w}_{s}\, \varepsilon_{s}^{ab}\,,
	\quad
	\varepsilon_{+}^{ab}=\frac{1}{\sqrt2}\left(e_\theta^a e_\theta^b-e_\phi^a e_\phi^b\right),
	\quad
	\varepsilon_{\times}^{ab}=\frac{1}{\sqrt2}\left(e_\theta^a e_\phi^b+e_\theta^b e_\phi^a\right)
\end{equation}
with $e_A^i = \partial_A n^i$.

Using the multipolar waveforms obtained in \cite{Georgoudis:2024pdz,Bini:2024rsy} and substituting into \eqref{eq:JijMultipoles} we find the following result in the PN limit,
\begin{equation}\label{eq:boldJ12}
	\begin{split}
		\boldsymbol{\boldsymbol{J}}_{xy}
		&=
		\frac{G^3 m^4 \pi}{b^2}
		\nu^2
		\left[
		\frac{28}{5}
		+
		\left(
		\frac{1679}{420}
		-
		\frac{79}{15}\,\nu
		\right)
		p_\infty^2
		+
\mathcal{O}(p_\infty^4)
		\right]
		\\
		&+
		\frac{G^4 m^5}{b^3 p^2_\infty}
		\nu^2
		\left[
		\frac{224}{5}
		+
		\left(
		\frac{12032}{105}
		-
		\frac{22832}{315}\,\nu
		\right)
		p_\infty^2
		+
\mathcal{O}(p_\infty^4)
		\right]
		\\
		&+
		\frac{G^4 m^5}{b^3}\,p_\infty
		\nu^2
		\left[
		\frac{448}{5}
		+
		\left(
		\frac{1184}{21}
		-
		\frac{45664}{315}\,\nu
		\right)
		p_\infty^2
		+
\mathcal{O}(p_\infty^4)
		\right].
	\end{split}
\end{equation}

We now consider the space-time components $\boldsymbol{J}_{i0}$, which are associated to the boost charge or mass dipole.
Owing again to the simplifying choices \eqref{eq:lambdadecomp}, \eqref{eq:Pimunusimplified}, which grant $\tilde{w}_{0 \mu}=0$, we have
\begin{equation}
	\boldsymbol{J}_{i0} = - \frac{1}{2i} \int_k 
	\left(
	\tilde{w}_{ab}^{\ast}(k) k^{\phantom{\mu}}_{[0} \frac{\partial}{\partial k^{i]}} \tilde{w}_{ab}(k)
	-
	\tilde{w}_{ab}(k) k^{\phantom{\mu}}_{[0}  \frac{\partial}{\partial k^{i]}} \tilde{w}^{\ast}_{ab}(k)
	\right).
\end{equation}
By means of Eqs.~\eqref{eq:changeofv} and following, we can recast $\boldsymbol{J}_{i0}$ as follows,
\begin{align}\label{eq:J0i_twolines}
		\kappa^2 \boldsymbol{J}_{i0} 
		&= G \int_0^\infty \frac{d\omega}{i\pi}\oint \frac{d\Omega}{4\pi}
		\left(
		\tilde{w}_{ab}^{\ast} \omega \partial_\omega \tilde{w}_{ab}
		-
		\tilde{w}_{ab} \omega \partial_\omega \tilde{w}^{\ast}_{ab}
		\right) 
		\omega\,
		n_i
		+
		\kappa^2 \boldsymbol{M}_i\,,
		\\
		\label{eq:Mi_twolines}
		\kappa^2 \boldsymbol{M}_i
		&
		=
		G \int_0^\infty \frac{d\omega}{i\pi}\oint \frac{d\Omega}{4\pi}
		\left(
		\tilde{w}_{ab}^{\ast} \partial_A \tilde{w}_{ab}
		-
		\tilde{w}_{ab} \partial_A \tilde{w}^{\ast}_{ab}
		\right)
		\omega\,
		\gamma^{AB} \partial_B n_i\,.
\end{align}
Explicitly,
\begin{subequations}
	\begin{align}
		\gamma^{AB}\partial_B n_x \partial_A &= \cos\theta \, \cos\phi\, \partial_\theta - \frac{\sin\phi}{\sin\theta}\, \partial_\phi \,,\\
		\gamma^{AB}\partial_B n_y \partial_A &= \cos\theta \, \sin\phi\, \partial_\theta + \frac{\cos\phi}{\sin\theta}\, \partial_\phi \,, \\
		\gamma^{AB}\partial_B n_z \partial_A &= -\sin\theta\, \partial_\theta\,.
	\end{align}
\end{subequations}

Using \eqref{eq:dotf_g} and \eqref{eq:tdotf_dotf} and taking into account the factors in \eqref{eq:waveform}, Eq.~\eqref{eq:J0i_twolines} can be recast as the integral over time of the following instantaneous flux in time domain
\begin{align}
	\label{eq:dBoldJi0}
	\frac{d \boldsymbol{J}_{i0} }{dt}
	&= t \left[ \frac{1}{32\pi G} \oint d\Omega\,
	\dot w^{ab} \dot w^{ab} n_i
	\right]
	+
	\frac{d\boldsymbol{M}_i}{dt}\,,
	\\
	\label{eq:dBoldMi0}
	\frac{d\boldsymbol{M}_i}{dt}
	&=
	- \frac{1}{32\pi G} \oint d\Omega
	(\dot w^{ab} \partial_A w^{ab}) \gamma^{AB} \partial_B
	n_i\,.
\end{align}
Recognizing the quantity within square brackets in Eq.~\eqref{eq:dBoldJi0} as the flux of radiated spatial momentum \eqref{eq:time-domain-flux-P}, this can be written equivalently as
\begin{equation}
	\frac{d \boldsymbol{J}_{i0} }{dt}
	-t\, \frac{d P_{i} }{dt}
	=
	\frac{d\boldsymbol{M}_i}{dt}
	\,.
\end{equation}
This equation defines a quantity, $\boldsymbol{M}_i$, after subtracting from $\boldsymbol{J}_{i0}$ the drift induced on the position of the center of mass due to the presence of a nontrivial recoil, $P_i$.
Recasting \eqref{eq:J0i_twolines} in terms of multipoles, one obtains the equivalent expression (let us recall the definition of $\overset{\leftrightarrow}{\partial}$ in \eqref{eq:leftrightarrow})
\begin{align}
	\begin{split}
		\frac{d\boldsymbol{J}_{i0}}{d\omega} 
		&=  
		\frac{2G\omega^2}{\pi} \sum_{\ell}^{\infty}\operatorname{Im}
		\Bigg[ 
		\frac{(\ell+2)(\ell+3) \boldsymbol{\mathrm{U}}_{i L}^* \overset{\leftrightarrow}{\partial}_\omega \boldsymbol{\mathrm{U}}_L}{\ell(\ell+1) !(2 \ell+3) ! !} +\frac{4(\ell+3) \boldsymbol{\mathrm{V}}^*_{i L} \overset{\leftrightarrow}{\partial}_\omega \boldsymbol{\mathrm{V}}_L }{(\ell+1) !(2 \ell+3) ! !} 
		\\
		&+ \frac{4(\ell+2) \epsilon_{i a b} \boldsymbol{\mathrm{U}}^*_{a L-1} \overset{\leftrightarrow}{\partial}_\omega \boldsymbol{\mathrm{V}}_{b L-1}}{(\ell-1)(\ell+1) !(2 \ell+1) ! !}  \Bigg]
		+
		\frac{d\boldsymbol{M}_i}{d\omega}\,
		\end{split}
		\\
		\frac{d\boldsymbol{M}_i}{d\omega}
		&=
		-\frac{2G\omega}{\pi} \operatorname{Im} \sum_\ell 
		\Bigg[
		\frac{(\ell+2)(\ell+3)}{\ell \ell! (2\ell+3)!!}\,
		\boldsymbol{\mathrm{U}}^\ast_{iL}
		\boldsymbol{\mathrm{U}}_L
		+
		\frac{4(\ell+3)}{\ell! (2\ell+3)!!}\,
		\boldsymbol{\mathrm{V}}^\ast_{iL}
		\boldsymbol{\mathrm{V}}_L
		\Bigg].
\end{align}

In order to apply \eqref{eq:J0i_twolines} to the waveforms obtained in \cite{Georgoudis:2024pdz,Bini:2024rsy}, we need to recall that 
\begin{equation}\label{eq:ambiguitymuIR}
	\tilde{w}_{ab} = \left(
	\frac{\omega}{\mu_\text{IR}}
	\right)^{2i G E\omega}
	\tilde{w}_{ab}^\text{reg}
\end{equation}
where the overall phase factor involves an arbitrary scale $\mu_{\text{IR}}$ that is left behind by the exponentiation of infrared divergences and amounts to an ambiguity in the definition of the origin of retarded time, while $\tilde{w}_{ab}^\text{reg}$ is $\mu_\text{IR}$-independent. Such a phase drops out in the calculation of the spectral rate \eqref{eq:spectralemissionrate} \cite{Georgoudis:2023eke}, which determines $P^\mu$, and the same happens in the calculation of $\boldsymbol{J}_{ij}$ \eqref{eq:Jij_twolines} because it is angle-independent in the center-of-mass frame.
Conversely, inserting \eqref{eq:ambiguitymuIR} into \eqref{eq:J0i_twolines} exposes the following dependence on the arbitrary scale in $\boldsymbol{J}_{i0}$,
\begin{align}\label{eq:J0i_threelines}
		\boldsymbol{J}_{i0} 
		&= 
		2GE \int_k \rho(k)
		\left(
		\log\frac{\omega}{\mu_{\text{IR}}}
		+
		1
		\right)
		\omega
		n_i
		+
		\boldsymbol{J}_{i0} ^\text{reg}\,,
		\\
		\label{eq:Ji0reg}
		\kappa^2 \boldsymbol{J}_{i0}^\text{reg}
		&=
		G \int_0^\infty \frac{d\omega}{i\pi}\oint \frac{d\Omega}{4\pi}
		\left(
		\tilde{w}^{\text{reg}\ast}_{ab} \omega \partial_\omega \tilde{w}^\text{reg}_{ab}
		-
		\tilde{w}^{\text{reg}}_{ab} \omega \partial_\omega \tilde{w}^{\text{reg}\ast}_{ab}
		\right) 
		\omega\,
		n_i
		+
		\kappa^2 \boldsymbol{M}_i
\end{align}
and, recognizing the radiated momentum \eqref{eq:Pioneline}, 
\begin{equation}\label{eq:muIRdmuIR}
	\mu_\text{IR}  \partial_{\mu_{\text{IR}}}
	\boldsymbol{J}_{i0} 
	= 
	-
2GE\, P_i\,.
\end{equation}
Thanks to the extra power of $G$ and to the fact that $P_x=\mathcal{O}(G^4)$, we thus see that $\boldsymbol{J}_{x0} $ is unaffected by the running scale at $\mathcal{O}(G^4)$. More generally, this running logarithm at $\mathcal{O}(G^{n+1})$ will drop out when focusing on the component orthogonal to $P_i$ at $\mathcal{O}(G^n)$.
Note also that, being proportional to the recoil and antisymmetric under $1\leftrightarrow2$, the ambiguity disappears for the equal-mass case, where $\boldsymbol{J}_{i0}=0$.
Finally, the $\mu_\text{IR}$-dependence also drops out from $\boldsymbol{M}_i$ which only involves angular derivatives, see \eqref{eq:Mi_twolines}. This quantity, which properly subtracts the drift induced by the presence of a nontrivial $P_i$, is thus completely unambiguous and provides a well-defined notion of mass-dipole loss in the center-of-mass frame. 

Let us now calculate the first few nonzero components of $\boldsymbol{J}_{i0}$ in the PN limit.
We thus find the radiative contributions 
\begin{equation}\label{eq:boldJ10}
	\begin{split}
		\boldsymbol{J}_{x0}
		&=
		\frac{G^3 \pi m^4}{b^2}
		p_\infty
		\nu^2	\sqrt{1-4\nu}
		\left[
		\frac{121}{30}
		+
		\left(
		\frac{1007}{560}
		-
		\frac{13}{4}\,\nu
		\right)
		p_\infty^2
		+
		\mathcal{O}(p_\infty^4)
		\right]
		\\
		&+
		\frac{G^4 m^5}{b^3 p_\infty}
		\nu^2	\sqrt{1-4\nu}
		\left[
		\frac{13712}{315}
		+
		\left(
		\frac{94168}{1575}
		-
		\frac{13576}{315}\,\nu
		\right)
		p_\infty^2
		+
		\mathcal{O}(p_\infty^4)
		\right]
		\\
		&+
		\frac{G^4 m^5}{b^3}\,p_\infty^2
		\nu^2	\sqrt{1-4\nu}
		\left[
		\frac{27424}{315}
		+
		\left(
		\frac{3296}{225}
		-
		\frac{27152}{315}\,\nu
		\right)
		p_\infty^2
		+
		\mathcal{O}(p_\infty^4)
		\right]\\
		&+\mathcal{O}(G^5)
	\end{split}
\end{equation}
and
\begin{equation}\label{eq:boldJ20}
	\begin{split}
	\boldsymbol{J}_{y0}^\text{reg}
	&=
	\frac{G^4  \pi m^5}{b^3}\,p_\infty^2
	\nu^2	\sqrt{1-4\nu}
	\left[
	\frac{36169}{3600}
	+
	\left(
	\frac{137}{5040}
	-
	\frac{15781}{7560}\,\nu
	\right)
	p_\infty^2
	+
	\mathcal{O}(p_\infty^4)
	\right]\\
	&
	+\mathcal{O}(G^5)\;.
	\end{split}
\end{equation}
Note that \eqref{eq:boldJ20} is sensitive both to the $\epsilon/\epsilon$ terms \emph{and} to the supertranslation frame discussed in \cite{Georgoudis:2024pdz,Bini:2024rsy}, for which we adopt the ``intrinsic'' choice \cite{Veneziano:2022zwh} which is more common in the PN literature.
As a cross-check, we explicitly verified that the $\mathcal{O}(G^3)$ contributions to \eqref{eq:boldJ12}, \eqref{eq:boldJ10} match those obtained by means of the covariant formula \eqref{eq:JmunuGaugeInv} and evaluated via reverse unitarity for generic velocities, see Eq.~(8.141) of \cite{DiVecchia:2023frv}.

Let us also remark that, for the scattering of scalar objects, any component involving one index in the $z$ direction necessarily vanishes, $\boldsymbol{J}_{\alpha z}=\zeta^\beta \boldsymbol{J}_{\alpha\beta}=0$, as a consequence of the fact that the integrand entering the covariant expression \eqref{eq:PmuGaugeInv} only depends on the external vectors $b_e^\mu$, $\tilde{u}_1^\mu$, $\tilde{u}_2^\mu$ which are all orthogonal to $\zeta^\mu$. The presence of misaligned spins leads instead to nontrivial $\boldsymbol{J}_{\alpha z}$ components \cite{Heissenberg:2023uvo,Jakobsen:2023hig}. 

For the cutoff-independent quantity $\boldsymbol{M}_i$, we find instead
\begin{equation}\label{eq:boldM1}
	\begin{split}
		\boldsymbol{M}_{x}
		&=
		\frac{G^3 \pi m^4}{b^2}
		p_\infty
		\nu^2	\sqrt{1-4\nu}
		\left[
		\frac{391}{105}
		+
		\left(
		\frac{269}{140}
		-
		\frac{319}{105}\,\nu
		\right)
		p_\infty^2
		+
		\mathcal{O}(p_\infty^4)
		\right]
		\\
		&+
		\frac{G^4 m^5}{b^3 p_\infty}
		\nu^2	\sqrt{1-4\nu}
		\left[
		\frac{352}{9}
		+
		\left(
		\frac{13904}{225}
		-
		\frac{368}{9}\,\nu
		\right)
		p_\infty^2
		+
		\mathcal{O}(p_\infty^4)
		\right]
		\\
		&+
		\frac{G^4 m^5}{b^3}\,p_\infty^2
		\nu^2	\sqrt{1-4\nu}
		\left[
		\frac{704}{9}
		+
		\left(
		\frac{1600}{63}
		-
		\frac{736}{9}\,\nu
		\right)
		p_\infty^2
		+
		\mathcal{O}(p_\infty^4)
		\right]\\
		&+\mathcal{O}(G^5)
	\end{split}
\end{equation}
and
\begin{equation}\label{eq:boldM2}
	\begin{split}
		\boldsymbol{M}_{y}
		&=
		\frac{G^4  \pi m^5}{b^3}\,p_\infty^2
		\nu^2	\sqrt{1-4\nu}
		\left[
		\frac{63}{10}
		+
		\left(
		-
		\frac{1021}{525}
		-
		\frac{2323}{700}\,\nu
		\right)
		p_\infty^2
		+
		\mathcal{O}(p_\infty^4)
		\right]\\
		&
		+\mathcal{O}(G^5)\;.
	\end{split}
\end{equation}

\subsection{Static contribution}
\label{ssec:calJ}

Let us now turn to the static part of \eqref{eq:JmunuGaugeInv}, starting from the contribution obtained from~\eqref{eq:opF1} which takes into account the gravitons emitted by the massive states,
\begin{equation}\label{eq:JmunuGaugeInvF}
	i \mathcal{J}_{\alpha\beta} =
	\int_k
	\left[
	\left(
	\eta^{\mu\rho}
	\eta^{\nu\sigma}
	-
	\tfrac{1}{D-2}\,
	\eta^{\mu\nu}
	\eta^{\rho\sigma}
	\right)
	{F}_{\mu\nu}^{\ast}
	k_{[\alpha}^{\phantom{\mu}}
	\frac{\overset{\leftrightarrow}{\partial} {F}_{\rho\sigma}}{\partial k^{\beta]}}
	+2
	\eta^{\mu\nu}
	F_{\mu[\alpha}^{\ast}
	F_{\beta]\nu}^{\phantom{\ast}}
	\right].
\end{equation}
This can be evaluated in general by substituting $F^{\mu\nu}$ given by \eqref{eq:Fmunu} and yields \cite{DiVecchia:2022owy}
\begin{subequations}
	\label{eq:JstaticLin}
	\begin{align}
	\mathcal{J}^{\alpha \beta}
	&=
	\frac{G}{2} 
	\sum_{a, b}
	c(\sigma_{ab})
	\left(\eta_{a}-\eta_{b}\right) p_{a}^{[\alpha} p_{b}^{\beta]}\,,
	\\
	c(\sigma_{ab})
	&=
	-
	\left[
	\left(
	\frac{\sigma^2_{ab}-\tfrac32}{\sigma^2_{ab}-1}
	\right)
	\frac{\sigma_{ab} \operatorname{arccosh}\sigma_{ab}}{\sqrt{\sigma^2_{ab}-1}}
	+\frac{\sigma^2_{ab}-\tfrac12}{\sigma^2_{ab}-1}
	\right],
	\end{align}
\end{subequations}
where $\eta_a$ is $+1$ (or $-1$) when the $a$ is outgoing (incoming), see Eq.~\eqref{eq:etaadef}, and
\begin{equation}
	\label{eq:sigmanm}
	\sigma_{ab} =  -\eta_a \eta_b\,\frac{p_a\cdot p_b}{m_a m_b}\,.
\end{equation}
For the scattering of scalars, the entire hyperbolic motion happens on the $xy$ plane, so that $\mathcal{J}_{\alpha z}=0$ in \eqref{eq:JstaticLin} vanishes identically.

Starting from \eqref{eq:JstaticLin}, we can substitute 
\begin{equation}\label{eq:G3accurate}
	\begin{split}
		p_1^\mu &= - \tilde m_1 \tilde u_1^\mu + \frac{1}{2}\,Q_1^\mu\,,\qquad
		p_4^\mu = \tilde m_1 \tilde u_1^\mu + \frac{1}{2}\,Q_1^\mu\,,\\
		p_2^\mu &= - \tilde m_2 \tilde u_2^\mu + \frac{1}{2}\,Q_2^\mu\,,\qquad
		p_3^\mu = \tilde m_2 \tilde u_2^\mu + \frac{1}{2}\,Q_2^\mu\,,
	\end{split}
\end{equation}
where $Q_1^\mu$, $Q_2^\mu$ are the full classical impulses that we will use up to at most $\mathcal O(G^3)$, and, at this order, \cite{Herrmann:2021lqe,Herrmann:2021tct}
\begin{equation}\label{eq:Q1Q2P}
	-\frac{Q_1\cdot b_e}{b_e}
	=
	+\frac{Q_2\cdot b_e}{b_e}
	=
	Q(b)\,,
	\qquad
	Q_1^\mu + Q_2^\mu = - {P}^\mu
\end{equation}
with $Q(b)$ explicitly given in \eqref{eq:Q3PM} below and $P^\mu$ the total emitted energy and momentum discussed in Section~\ref{sec:P}.
We specify the center-of-mass frame by choosing the reference vectors again as in \eqref{eq:refvec}. In this way, the chosen boost-frame satisfies $\tilde m_{1} \tilde{u}_1^i + \tilde m_{2} \tilde {u}_2^i =0$ in terms of the average or ``eikonal'' momenta, while $p_1^i + p_2^i = -\tfrac12\,P^i =\mathcal{O}(G^3)$ due to recoil. An alternative possibility would be to specify this frame by enforcing the exact condition $p_1^i+p_2^i=0$ in terms of the incoming momenta. Note that this is related to our frame choice by a boost that differs from the identity by $\mathcal{O}(G^3)$ terms; since as we shall see shortly $\mathcal{J}^{\alpha\beta}=\mathcal{O}(G^2)$ to leading order, this boost would not change the explicit results presented below up to and including $\mathcal{O}(G^4)$. The translation-frame is instead immaterial here, because the static part of the angular momentum does not transform under translations (being localized at $\omega=0$).

It is then possible to write down explicitly the first few orders in the PM expansion of~\eqref{eq:JstaticLin}. One needs to relate the initial and the final momenta by using~\eqref{eq:G3accurate} and expand for small deflections. In order to facilitate the comparison with the literature, while retaining the coordinate system aligned as in~\eqref{eq:refvec}, we give the results in terms of $m_i$ and $b$, rather than $b_e$ and $\tilde{m}_i$ (see also footnote~\ref{footnote:bebJ}). It is convenient to introduce the function
\begin{equation}\label{eq:isigma}
        \mathcal{I}(\sigma)
        =
        -\frac{16}{3}+\frac{2\sigma^2}{\sigma^2-1}+\frac{2\sigma(2\sigma^2-3) \operatorname{arccosh}\sigma}{(\sigma^2-1)^{3/2}}\,,
\end{equation}
so we can write the component $\mathcal{J}_{xy}$ up to and including $\mathcal{O}(G^4)$ as follows
\begin{align} \label{eq:calJ4PM} 
                \mathcal{J}_{xy}
                &=
                G\, p\,Q(b)\, \mathcal{I}(\sigma) 
                \\ \nonumber
                &
                -
                \frac{G Q(b)^3}{E\sqrt{\sigma^2-1}}
                \left[
                \frac{E^2}{8 m_1 m_2}
                \left(
                        \mathcal{I}(\sigma)+\frac{32}{5}\,(\sigma^2-1)
                        \right)
                        +
                        \frac{1}{2}\left(
                        \mathcal{I}'(\sigma)-\frac{16}{5}\sigma
                        \right)
                        (\sigma^2-1)
                \right]
                \\ \nonumber
                &+\mathcal{O}(G^5)\,,
\end{align}
where $Q(b)$ is the component of $Q_1^\mu$ along $-b_e^\mu$ (see Eq.~\eqref{eq:Q1Q2P}) expressed as a function of the initial impact parameter $b$, \cite{Bern:2019nnu,Bern:2019crd,Damour:2020tta}
\begin{align}\label{eq:Q3PM}
        &Q(b) 
        = \frac{4 G m_1 m_2 \left(\sigma ^2-\frac{1}{2}\right)}{b \sqrt{\sigma ^2-1}}
        +
        \frac{3 \pi  G^2 m_1 m_2 \left(m_1+m_2\right) \left(5 \sigma ^2-1\right)}{4 b^2 \sqrt{\sigma ^2-1}}
        +\frac{8 G^3 m_1^2 m_2^2}{b^3}
        \\
        \nonumber
        &\times \!\left(\!\frac{\left(-4 \sigma ^4+12 \sigma ^2+3\right) \operatorname{arccosh} \sigma }{\sigma
                        ^2-1}+\frac{E^2 \left(12 \sigma ^4-10 \sigma ^2+1\right)}{2 m_1 m_2 \left(\sigma
                        ^2-1\right)^{3/2}}-\frac{\sigma  \left(14 \sigma ^2+25\right)}{3 \sqrt{\sigma ^2-1}}\!\right)
         \\
         \nonumber
        &+\frac{G}{2b}\left(
        \frac{4 G m_1 m_2 \left(\sigma ^2-\frac{1}{2}\right)}{b \sqrt{\sigma ^2-1}}
        \right)^2 \mathcal{I}(\sigma)
        -\frac{G^3 m_1 m_2 E^2 \left(2 \sigma ^2-1\right)^3}{b^3 \left(\sigma ^2-1\right)^{5/2}}
        +
        \mathcal{O}(G^4)\;.
\end{align}
We can then make contact with the PN expressions by further expanding the PM result for small $p_\infty$. At the first  few PN orders, one obtains
\begin{equation}\label{eq:calJ12}
	\begin{split}
		\mathcal{J}_{xy}
		&=
		\frac{G^2 m^3}{b}
		p_\infty^2
		\nu^2
		\left[
		\frac{16}{5}
		+
		\left(
		\frac{176}{35}
		-
		\frac{8}{5}\,\nu
		\right)
		p_\infty^2
		+
\mathcal{O}(p_\infty^4)
		\right]
		\\
		&+
		\frac{G^3 m^4 \pi}{b^2}		
		\nu^2
		\left[
		\frac{24}{5} p_\infty^2
		+
\mathcal{O}(p_\infty^4)
		\right]
		\\
		&+
		\frac{G^4 m^5}{b^3p_\infty^2}
		\nu^2
		\left[
		-\frac{48}{5}
		+
		\left(
		-\frac{1296}{35}
		+
		\frac{216}{35}\,\nu
		\right)
		p_\infty^2
		+
		\mathcal{O}(p_\infty^4)
		\right]
		\\
		&+
		\frac{G^4 m^5}{b^3}\,p_\infty
		\nu^3
		\left[
		\frac{128}{25} p_\infty^2
		+
		\mathcal{O}(p_\infty^4)
              \right]\\
              & +\mathcal{O}(G^5)\,.
	\end{split}
\end{equation}

By following the same approach, one can evaluate also the time components of $\mathcal{J}_{\alpha\beta}$ relevant for the mass dipole of the radiation. Up to order $\mathcal{O}(G^4)$ we have
\begin{equation}
        \mathcal{J}_{x0}
        =
        -\frac{G(m_1^2-m_2^2)}{2E}
        \left[
        Q(b) \mathcal{I}(\sigma)
        +
        \frac{Q(b)^3 \mathcal{G}(\sigma)}{m_1 m_2 (\sigma^2-1)}
        \right]
        +
        \mathcal{O}(G^5),
\end{equation}
where
\begin{equation}
        \mathcal{G}(\sigma) 
        =
        -\frac{3\operatorname{arccosh}\sigma}{(\sigma^2-1)^{3/2}} 
        +
        \frac{8 \sigma ^5-26 \sigma ^3+33 \sigma}{5(\sigma^2-1)}\,
\end{equation}
and for the $y0$ component
\begin{equation}
        \mathcal{J}_{y0}
        =
        \frac{GE \, E_\text{rad}}{\sqrt{\sigma^2-1}}\sqrt{1-4\nu}\,\left(
        \mathcal{H}(\sigma)
        -\frac{\sigma}{2}\,\mathcal{I}(\sigma)
        \right)
        +
        \mathcal{O}(G^5),
\end{equation}
where
\begin{equation}
        \mathcal{H}(\sigma) 
        =
        -\frac{3\operatorname{arccosh}\sigma}{(\sigma^2-1)^{3/2}} 
        +
        \frac{-17 \sigma ^3+11 \sigma ^2+26 \sigma -11}{3(\sigma^2-1)}
\end{equation}
and $E_\text{rad}$ can be found in Eq.~(10) of Ref.~\cite{Herrmann:2021lqe}.
Further expanding for small $p_\infty$ we have
\begin{equation}\label{eq:calJ10}
	\begin{split}
		\mathcal{J}_{x0}
		&=
		\frac{G^2 m^3}{b}
		p_\infty
		\nu	\sqrt{1-4\nu}
		\left[
		-
		\frac{8}{5}
		+
		\left(
		-
		\frac{88}{35}
		+
		\frac{4}{5}\,\nu
		\right)
		p_\infty^2
		+
		\mathcal{O}(p_\infty^4)
		\right]
		\\
		&+
		\frac{G^3 m^4 \pi}{b^2}
		p_\infty
		\nu	\sqrt{1-4\nu}
		\left[
		-
		\frac{12}{5}
		+
		\left(
		-
		\frac{69}{35}
		+
		\frac{6}{5}\,\nu
		\right)
		p_\infty^2
		+
		\mathcal{O}(p_\infty^4)
		\right]
		\\
		&+
		\frac{G^4 m^5}{b^3 p_\infty^3}
		\nu	\sqrt{1-4\nu}
		\left[
		\frac{4}{5}
		+\left(
		-\frac{36}{7}
		-\frac{178}{35}\,\nu
		\right)
		p_\infty^2
		\right.
		\\
		&\
		\left.
		+
		\left(
		-\frac{10376}{315}
		-
		\frac{4337}{210}\,\nu
		+
		\frac{37}{14}\,\nu^2
		\right)
		p_\infty^4
		+
		\mathcal{O}(p_\infty^6)
		\right]
		\\
		&+
		\frac{G^4 m^5}{b^3}\,p_\infty^2
		\nu^2	\sqrt{1-4\nu}
		\left[
		-
		\frac{64}{25}
		+
		\left(
		-
		\frac{1408}{175}
		+
		\frac{32}{25}\,\nu
		\right)
		p_\infty^2
		+
		\mathcal{O}(p_\infty^4)
              \right]\\
              &+\mathcal{O}(G^5),
	\end{split}
\end{equation}
while $\mathcal{J}_{z0}=0$,
\begin{equation}\label{eq:calJ20}
	\mathcal{J}_{y0}
	=
	\frac{G^4 \pi m^5}{b^3}\,p_\infty^2
	\nu^2	\sqrt{1-4\nu}
	\left[
	-
	\frac{4699}{450}
	-
	\frac{15983}{3600}\,
	p_\infty^2
	+
	\mathcal{O}(p_\infty^4)
	\right]+\mathcal{O}(G^5).
\end{equation}

Let us note that the TT-projected formula analogous to the standard one \eqref{eq:JmunuTT}, 
\begin{equation}\label{eq:JmunuTTstat}
	i \mathcal{J}^\text{TT}_{\alpha\beta} =
	\int_k
	\left[
	\tfrac{1}{2}
	\left(
	f_{\mu\nu}^{\ast}
	k_{[\alpha}^{\phantom{\mu}}
	\frac{\partial f^{\mu\nu}}{\partial k^{\beta]}}
	-
	f^{\mu\nu}
	k_{[\alpha}^{\phantom{\mu}}
	\frac{\partial f^\ast_{\mu\nu}}{\partial k^{\beta]}}
	\right)
	+2
	f_{\mu[\alpha}^{\ast}
	f_{\beta]}^{\mu}
	\right],
\end{equation}
with 
\begin{equation}
f^{\mu\nu} = \Pi_{\mu\nu,\rho\sigma} F^{\mu\nu}\,,
\end{equation}
and the manifestly covariant one \eqref{eq:JmunuGaugeInvF} are not equivalent when it comes to the static contributions \cite{Jakobsen:2021smu,Mougiakakos:2021ckm,Manohar:2022dea,Riva:2023xxm}. 
In particular, \eqref{eq:JmunuTTstat} explicitly depends on the reference vector $\lambda^\mu$ defining the TT projection \eqref{eq:TTprojector}, contrary to what happens for the radiative case \eqref{eq:JmunuTT}.
As explained in Ref.~\cite{Riva:2023xxm}, the reason for this discrepancy is that $\mathcal{J}^\text{TT}_{\alpha\beta}$ suppresses the contribution due to the Coulombic field and thus does not admit a covariant uplift.  More in detail, Ref.~\cite{Riva:2023xxm} showed that $\mathcal{J}_{\alpha\beta}$ can be expressed  up to $\mathcal O(G^4)$ corrections as an integral involving the function
\begin{equation}\label{eq:DeltaS}
	\Delta S = 2G \sum_{a} p_a\cdot n \log\left(-\eta_a\,\frac{p_a\cdot n}{m_a}\right),
\end{equation}
see Eq.~(3.25) of that reference, while $\mathcal{J}^\text{TT}_{\alpha\beta}$ given by \eqref{eq:JmunuTTstat}  is equivalent to the same integral where $\Delta S$ is replaced with its projection on $\ell\ge2$ spherical harmonics. This operation amounts to dropping the contribution due to the Coulombic static field, and spoils the Lorentz covariance of the result.

The inequivalence between \eqref{eq:JmunuGaugeInvF} and \eqref{eq:JmunuTTstat} can be traced back to the fact that, letting $k^\mu = \omega \, n^\mu$, and including a factor of $\frac12$ to avoid double counting \cite{Manohar:2022dea},
\begin{equation}\label{eq:nontransv}
	n_\mu F^{\mu\nu} = i\pi \sqrt{8\pi G} \, \sum_{a\in\text{in}}p_a^\nu \delta(\omega)\neq 0
\end{equation} 
(although $k_\mu F^{\mu\nu} = 0$ as a distribution by  the identity $\omega\delta(\omega)=0$).
One can check that the distinction between \eqref{eq:JmunuGaugeInvF} and \eqref{eq:JmunuTTstat} is irrelevant for the spatial angular momentum evaluated in the center-of-mass frame, owing to the form of $n_\mu F^{\mu\nu}$ in \eqref{eq:nontransv},
\begin{equation}\label{eq:calJijTT}
	\mathcal J_{ij}^\text{TT} = \mathcal J_{ij}\,.
\end{equation} 
In particular, this equality relies on the fact that $n_\mu F^{\mu\nu}$ is angle independent and that $n_\mu F^{\mu i}$ vanishes for $i = 1,2,3$ in the center-of-mass frame.
Instead, the distinction between \eqref{eq:JmunuGaugeInvF} and \eqref{eq:JmunuTTstat} is important for the mass dipole in this frame, for which we find
(see Appendix~\ref{app:upliftTT} for more details)
\begin{equation}\label{eq:calJi0TT}
	\mathcal{M}_i
	=
	\mathcal{J}^{\text{TT}}_{i0}
	=
	\mathcal{J}_{i0}
	-
	2GE
	\sum_{a}
	c(\sigma_a)\,
	p_{a}^i\,,
\end{equation}
where $c(\sigma_a)$ is the same function appearing in \eqref{eq:JstaticLin} but now evaluated at the following argument,
\begin{equation}
	\sigma_a = \frac{E_a}{m_a}
\end{equation}
with $E_a = \eta_a\,p_a^0$  the energy of the $a$th state in the center-of-mass frame.
The complete tensor \eqref{eq:JmunuTTstat} can be thus written as follows\footnote{While still formally covariant, the result \eqref{eq:calJalphabetaFinal} depends on the properties of the reference vector $\lambda^\mu$ chosen in \eqref{eq:lambdadecomp}, e.g.~on the fact that $n_\mu F^{\mu\nu} \lambda_\nu$ is angle-independent.}
\begin{equation}\label{eq:calJalphabetaFinal}
	\mathcal{J}^{\text{TT}\alpha\beta} = \frac{G}{2}\sum_{a,b}c(\sigma_{ab})\left(
	\eta_a-\eta_b
	\right)
	p_a^{[\alpha}
	p_b^{\beta]}
	+ 2G \sum_{a} c(\sigma_a) \sum_{b\in\text{in}}p_b^{[\alpha}p_a^{\beta]}\,.
\end{equation}
Incidentally, we note that $\mathcal{J}^{\text{TT}\alpha\beta}$ in \eqref{eq:calJalphabetaFinal} admits the smooth massless limit\footnote{We evaluate this limit for formally elastic dynamics of the hard process.}
\begin{equation}
	\mathcal{J}^{\text{TT}\alpha\beta} 
	= - 2G \log\left(\frac{E^2}{Q^2}-1\right)
	(p_1-p_2)^{[\alpha} Q^{\beta]}\,,
\end{equation}
which can be taken as an indication that the \emph{mechanical} mass dipole moment should include the Coulombic contribution in order to obtain a well-defined ultrarelativistic limit.
Note also that
$\mathcal{M}_i \neq \mathcal J_{i0}$ already starting at $\mathcal{O}(G^2)$.
For instance, 
\begin{equation}
	\mathcal{M}_x
	=
	\mathcal{J}_{x0}
	+
	2GE Q(b) \left[
	c\left(
	\sigma_1
	\right)
	-
	c\left(
	\sigma_2
	\right)
	\right]
	+\mathcal{O}(G^5)\,,
\end{equation}
whose PN expansion reads
\begin{equation}\label{eq:calJ10TT}
	\begin{split}
		\mathcal{M}_x
		&=
		\frac{G^2 m^3}{b}
		p_\infty^3
		\nu^2	\sqrt{1-4\nu}
		\left[
		\frac{48}{35}
		+
		\left(
		\frac{568}{315}
		-
		\frac{88}{105}\,\nu
		\right)
		p_\infty^2
		+
		\mathcal{O}(p_\infty^4)
		\right]
		\\
		&+
		\frac{G^3 m^4 \pi}{b^2}
		p_\infty
		\nu^2	\sqrt{1-4\nu}
		\left[
		\frac{72}{35} \, p_\infty^2
		+
		\mathcal{O}(p_\infty^4)
		\right]
		\\
		&+
		\frac{G^4 m^5}{b^3 p_\infty}
		\nu^2	\sqrt{1-4\nu}
		\left[
		- \frac{216}{35}
		+\left(
		-\frac{8252}{315}
		+\frac{52}{21}\,\nu
		\right)
		p_\infty^2
		+
		\mathcal{O}(p_\infty^4)
		\right]
		\\
		&+
		\frac{G^4 m^5}{b^3}\,p_\infty^2
		\nu^2	\sqrt{1-4\nu}
		\left[
		\frac{384}{175}\,
		p_\infty^2
		+
		\mathcal{O}(p_\infty^4)
		\right]
		+\mathcal{O}(G^5)\,.
	\end{split}
\end{equation}
Finally, letting 
\begin{align}
	\alpha_0 &= 2 m_2^2 m_1^2 \left(\sigma ^3+2\right)+m_1 m_2 (m_1^2+m_2^2) (3 \sigma +1)+m_1^4+m_2^4\,,
	\\
	\begin{split}
	\alpha_1 &=
	-2 m_2 m_1^3 \sigma  \left(\sigma ^2+3\right)-3 m_2^2 m_1^2 \left(3 \sigma ^2+1\right)
	\\
	&+m_1^4 \left(-2 \sigma
	^4+3 \sigma ^2-3\right)-8 m_2^3 m_1 \sigma -2 m_2^4
	\end{split}
\end{align}
and $\alpha_2(m_1,m_2)=\alpha_1(m_2,m_1)$,
we find 
\begin{equation}
	\begin{split}
	\mathcal{M}_{y}
	&=
	\mathcal{J}_{y0}
	+
	\frac{GE \, E_\text{rad}}{m}
	\Bigg[
	\frac{2(m_1-m_2)\alpha_0}{m_1^2 m_2^2 (\sigma^2-1)^{3/2}}
	-\frac{\alpha_1 \operatorname{arccosh}\sigma_2}{m_1^3(\sigma^2-1)^2}
	+\frac{\alpha_2 \operatorname{arccosh}\sigma_1}{m_2^3(\sigma^2-1)^2}
	\Bigg]
	\\
	&+
	\mathcal{O}(G^5),
	\end{split}
\end{equation}
whose PN expansion gives
\begin{equation}\label{eq:calJ20TT}
	\mathcal{M}_y = \frac{G^4 \pi m^5 }{b^3}\,p_\infty^2\nu^2
	\sqrt{1-4\nu}\,\left[
	-\frac{74}{175}\,\nu\,p_\infty^2+\mathcal{O}(p_\infty^4)
	\right].
\end{equation}

The zero-frequency limit \eqref{eq:Fmunu} captures only linear contributions, i.e.~the  field sourced by the massive lines. Including nonlinear ones, that is soft/static fields generated by dynamically produced gravitons, amounts to replacing \eqref{eq:Fmunu} with~\eqref{eq:FmunuT}. Following~\cite{Christodoulou:1991cr,Wiseman:1991ss,Thorne:1992sdb}, we replace the sum over the hard gravitons in~\eqref{eq:FmunuT} by a phase-space integral weighted by the spectral emission rate $\rho(k)$ introduced in~\eqref{eq:spectralemissionrate},
\begin{equation}\label{eq:Ftot}
	F_\text{tot}^{\mu\nu}(\ell)
	=
	F^{\mu\nu}(\ell)
	+
	\delta F^{\mu\nu}(\ell)
	=
	\sum_{a_m} \frac{\sqrt{8\pi G}\,  p_{a_m}^\mu p_{a_m}^\nu}{p_{a_m}\cdot \ell-i0} + \int_k \rho(k) \, \frac{\sqrt{8\pi G}\,k^\mu k^\nu}{ k \cdot \ell-i0}\,.
\end{equation}
Notice that the last term has a collinear divergence in $D=4$ because of the denominator, which however disappears when taking the TT projection.

It is convenient to note that \eqref{eq:Ftot} is obtained from \eqref{eq:Fmunu} by formally extending the sum also to outgoing gravitons according to
\begin{equation}\label{eq:operation}
	\sum_a \mapsto \sum_{a_m} + \int_k \rho(k)\,.
\end{equation}
Applying this operation on \eqref{eq:JstaticLin} and noting that, as $m_b\to0$
\begin{equation}\label{eq:log+1}
	\left[
	\left(
	\frac{\sigma^2_{ab}-\tfrac32}{\sigma^2_{ab}-1}
	\right)
	\frac{\sigma_{ab}\operatorname{arccosh}\sigma_{ab}}{\sqrt{\sigma^2_{ab}-1}}
	+\frac{\sigma^2_{ab}-\tfrac12}{\sigma^2_{ab}-1}
	\right]
	\sim 
	\log\frac{-2\eta_a\eta_b p_a\cdot p_b}{m_a m_b} + 1 + o(m_b)\,,
\end{equation}
we find
\begin{equation}\label{eq:calJ+deltacalJ}
	\mathcal{J}_{\alpha \beta}
	\mapsto 
	\mathcal{J}^{\rm tot}_{\alpha \beta} = \mathcal{J}_{\alpha \beta}
	+ \delta\mathcal{J}_{\alpha\beta}
	+ \mathcal {O}(G^7)\,,
\end{equation}
where the first correction due to nonlinear memory is
\begin{equation}\label{eq:deltacalJ}
	\delta\mathcal{J}^{\alpha\beta} = 2G \int_k \rho(k) 
	\sum_{a\in\text{in}}
	p_a^{[\alpha} k^{\beta]}
	\log\frac{p_a\cdot k}{m_a \Lambda} 
\end{equation}
and $\Lambda$ is an energy scale introduced to regulate the collinear divergence.
In \eqref{eq:calJ+deltacalJ}, we used that $\rho(k)\sim\mathcal{O}(G^3)$ to leading order to neglect terms quadratic in the nonlinear memory effect.
Note that the dependence on this regulator enters \eqref{eq:deltacalJ} as follows,
\begin{equation}\label{eq:generalcutoffdependence}
	\Lambda \partial_\Lambda \delta \mathcal{J}^{\alpha\beta} = -2G (p_1+p_2)^{[\alpha} P^{\beta]}
\end{equation}
where we used \eqref{eq:PmuGaugeInv}.

Focusing first on the spatial components, we recall that $p_1^i + p_2^i =-\frac{1}{2}\,P^i$ in the center-of-mass frame we adopt and thus by \eqref{eq:generalcutoffdependence} the cutoff dependence drops out from these components,
\begin{equation}
	 \Lambda \partial_\Lambda \delta \mathcal{J}_{ij} = 0\,.
\end{equation}
This observation, combined with \eqref{eq:calJ+deltacalJ}, ensures that $\mathcal{J}^{\text{tot}}_{ij}$ is completely well defined up to and including $\mathcal{O}(G^6)$.
In terms of the basis vectors \eqref{eq:refvec}, we thus obtain\footnote{Here and in the following, we can use the $\mathcal{O}(G^2)$-accurate kinematics, up to $\mathcal{O}(G^6)$ corrections.} (we recall that $v_1^\mu$, $v_2^\mu$ are the initial velocities defined by \eqref{eq:velocityandsigma})
\begin{equation}
	\label{eq:deltaJxy}
	\delta\mathcal{J}_{xy} 
	=\frac{2Gp}{b_e} \int_k \rho(k)(b_e\cdot k)
	\log\frac{v_1\cdot k}{v_2 \cdot k}
	+\mathcal{O}(G^5)\,,
\end{equation}
where $p E=m_1 m_2 \sqrt{\sigma^2-1}$.
In this way, we see that the spatial angular momentum loss due to the nonlinear memory effect is actually independent of the arbitrary scale $\Lambda$. In fact, it vanishes at $\mathcal{O}(G^4)$ order because the leading $\mathcal{O}(G^3)$ spectral emission rate $\rho_0$ in \eqref{eq:rho0} is \emph{even} under $b_e\cdot k\mapsto - b_e\cdot k$ \eqref{eq:rho0isEVEN}, and thus the integrand resulting from \eqref{eq:deltaJxy} is \emph{odd}. We have explicitly cross-checked the cancellation ensured by this observation at the first few PN orders. 
This parity argument does not apply to the $\mathcal{O}(G^5)$ contribution, which is sensitive to the one-loop spectral emission rate, and is generically nonvanishing, while still independent of the arbitrary scale $\Lambda$.

However, the dependence on the regulator $\Lambda$ in \eqref{eq:deltacalJ} does generically survive in the mass-dipole loss,
\begin{equation}\label{eq:LambdadLambda}
	\Lambda \partial_\Lambda \delta\mathcal{J}_{i0} 
	=
	2GE\, {P}_{i} +\mathcal{O}(G^6)\,,
\end{equation} 
where ${P}^\alpha$ is the radiated energy-momentum \eqref{eq:PmuGaugeInv}. 
Comparing \eqref{eq:muIRdmuIR} with \eqref{eq:LambdadLambda}, we note that the $\log\Lambda$ ambiguity in $\delta\mathcal{J}_{y0}$ appears with coefficient opposite to the one of $\log\mu_\text{IR}$ in $\boldsymbol{J}_{y0}$. 

Let us note that, since the collinear divergence is absent in the TT projection of the nonlinear memory waveform $\delta F^{\mu\nu}$, the $\Lambda$-dependence  disappears entirely if we calculate its contribution to the angular momentum loss by generalizing \eqref{eq:JmunuTTstat} instead of \eqref{eq:JmunuGaugeInvF}. 
Indeed, applying the operation \eqref{eq:operation} to \eqref{eq:calJalphabetaFinal} instead of \eqref{eq:JstaticLin}, we find
\begin{equation}\label{eq:deltacalJTT}
	\delta\mathcal{J}^{\text{TT}\alpha\beta} = 2G \int_k \rho(k) 
	\sum_{a\in\text{in}}
	p_a^{[\alpha} k^{\beta]}
	\log\frac{p_a\cdot n}{m_a}\,,
\end{equation}
where $n^\mu = (1,\hat n)$ as defined in the center-of-mass frame.
As already remarked, \eqref{eq:deltacalJTT} leads to the same prediction as \eqref{eq:deltacalJ} for the $ij$ components, and in particular to  \eqref{eq:deltaJxy}.
This is because $n_\mu F^{\mu\nu}_\text{tot}=n_\mu F^{\mu\nu}$ is still as given in \eqref{eq:nontransv}, and thus \eqref{eq:calJijTT}  can be upgraded to the nonlinear case as well, $\delta \mathcal{J}^\text{TT}_{ij} = \delta \mathcal{J}_{ij}$.
Instead, letting again $\delta\mathcal{M}_i = \delta \mathcal{J}^\text{TT}_{i0}$, we find that 
\begin{equation}
		\delta\mathcal{J}_{i0} 
	=
	- 2GE \int_k \rho(k)\,\omega n_i \log\frac{\omega}{\Lambda}
	+
	\delta\mathcal{M}_{i}\,, 
\end{equation}
where it is clear that the entire integral involving $\log\omega$ will eventually cancel out against the one in \eqref{eq:J0i_threelines}.
More explicitly,
\begin{align}
	\label{eq:deltaJxtreg}
	\delta\mathcal{M}_{x} 
	&= -\frac{2G}{b} \int_k \rho(k) (b_e\cdot k)\sum_{a\in\text{in}}
	E_a 
	\log\frac{p_a\cdot n}{m_a}+\mathcal{O}(G^5) \,,
	\\
	\label{eq:deltaJytEXPL}
	\delta\mathcal{M}_{y}
	&=
	2G \int_k \rho(k) 
	\,
	\omega
	\left[
	p
	\log\frac{v_1\cdot k}{v_2\cdot k} 
	-
	n_y
	\sum_{a\in\text{in}}
	E_a 
	\log\frac{p_a\cdot n}{m_a} 
	\right]+\mathcal{O}(G^5).
\end{align}
The same parity argument discussed below \eqref{eq:deltaJxy} ensures that $\delta\mathcal{J}_{x0}$  and $\delta\mathcal{M}_{x}$
vanish altogether at order $\mathcal{O}(G^4)$.
Instead, using the explicit tree-level waveforms, we find
\begin{equation}\label{eq:regexpl}
	\delta\mathcal{M}_{y}
	=
	\frac{G^4 \pi m^5}{b^3}p_\infty^2
	\nu^2
	\sqrt{1-4\nu}
	\left[
	\frac{15781}{15120}\,\nu\,p_\infty^2 + \mathcal{O}(p_\infty^2)
	\right]
	+\mathcal{O}(G^5)\,.
\end{equation}

\subsection{Complete expressions}

Adding \eqref{eq:boldJ12} and \eqref{eq:calJ12}, we recover the following PN-expanded expression for the total angular momentum loss, which is aligned with the $z$-direction orthogonal to the scattering plane,
\begin{equation}\label{eq:Jxy}
	\begin{split}
		{J}_{xy}
		&=
		\frac{G^2 m^3}{b}
		p_\infty^2
		\nu^2
		\left[
		\frac{16}{5}
		+
		\left(
		\frac{176}{35}
		-
		\frac{8}{5}\,\nu
		\right)
		p_\infty^2
		+
		\mathcal{O}(p_\infty^4)
		\right]
		\\
		&+
		\frac{G^3 \pi m^4}{b^2}
		\nu^2
		\left[
		\frac{28}{5}
		+
		\left(
		\frac{739}{84}
		-
		\frac{79}{15}\,\nu
		\right)
		p_\infty^2
		+
		\mathcal{O}(p_\infty^4)
		\right]
		\\
		&+
		\frac{G^4 m^5}{b^3 p^2_\infty}
		\nu^2
		\left[
		\frac{176}{5}
		+
		\left(
		\frac{8144}{105}
		-
		\frac{2984}{45}\,\nu
		\right)
		p_\infty^2
		+
		\mathcal{O}(p_\infty^4)
		\right]
		\\
		&+
		\frac{G^4 m^5}{b^3}\,p_\infty
		\nu^2
		\left[
		\frac{448}{5}
		+
		\left(
		\frac{1184}{21}
		-
		\frac{220256}{1575}\,\nu
		\right)
		p_\infty^2
		+
		\mathcal{O}(p_\infty^4)
		\right]+\mathcal{O}(G^5).
	\end{split}
\end{equation}
The first two lines of \eqref{eq:Jxy} reproduce the small-velocity expansion of the $\mathcal{O}(G^2)$ and $\mathcal{O}(G^3)$ results in \cite{Damour:2020tta} and \cite{Manohar:2022dea,DiVecchia:2022piu}.
The last two lines of \eqref{eq:Jxy} are in perfect agreement with the 0PN, 1PN, 1.5PN and 2.5PN contributions at $\mathcal{O}(G^4)$ obtained in \cite{Bini:2021gat,Bini:2022enm}.

Moving to the mass dipole, the sum of \eqref{eq:boldJ10} and \eqref{eq:calJ10} yields instead
\begin{equation}
	\begin{split}
		{J}_{x0}
		&=
		\frac{G^2 m^3}{b}
		p_\infty
		\nu	\sqrt{1-4\nu}
		\left[
		-
		\frac{8}{5}
		+
		\left(
		-
		\frac{88}{35}
		+
		\frac{4}{5}\,\nu
		\right)
		p_\infty^2
		+
		\mathcal{O}(p_\infty^4)
		\right]
		\\
		&+
		\frac{G^3 m^4 \pi}{b^2}
		p_\infty
		\nu	\sqrt{1-4\nu}
		\Bigg[
		-
		\frac{12}{5}
		+
		\frac{121}{30}\,\nu
		\\
		&+
		\left(
		-
		\frac{69}{35}
		+
		\frac{1679}{560}\,\nu
		-
		\frac{13}{4}\,\nu^2
		\right)
		p_\infty^2
		+
		\mathcal{O}(p_\infty^4)
		\Bigg]
		\\
		&+
		\frac{G^4 m^5}{b^3 p_\infty^3}
		\nu	\sqrt{1-4\nu}
		\Bigg[
		\frac{4}{5}
		+
		\left(-\frac{36}{7}+\frac{346}{9}\,\nu\right) p_\infty^2
		\\
		&+
		\left(-\frac{10376}{315}+\frac{123281}{3150}\,\nu-\frac{3641}{90}\,\nu^2\right) p_\infty^4
		+
		\mathcal{O}(p_\infty^6)
		\Bigg]
		\\
		&+
		\frac{G^4 m^5}{b^3}\,p_\infty^2
		\nu^2	\sqrt{1-4\nu}
		\left[
		\frac{133088}{1575}
		+\left(\frac{416}{63}
		-\frac{133744}{1575}\,\nu\right) p_\infty^2
		+
		\mathcal{O}(p_\infty^4)
		\right],
		\\
		&+\mathcal{O}(G^5)\,.
	\end{split}
\end{equation}
which corresponds to a well-defined change in the mass-dipole component along the impact parameter $b_e^\mu$.
Finally, from the sum of \eqref{eq:boldJ20}, \eqref{eq:calJ20} and \eqref{eq:regexpl},
\begin{align}\label{eq:Jy0final}
	&J_{y0}
	=
	2GE
	P_y
	\left(\log\frac{\Lambda}{\mu_\text{IR}} + 1 \right)
	\\
	\nonumber
	&
	+\frac{G^4  \pi m^5}{b^3}\,p_\infty^2
	\nu^2	\sqrt{1-4\nu}
	\left[
	-\frac{1423}{3600}
	+
	\left(
	-\frac{27799}{6300}
	+
	\frac{15781}{15120} 
	\,\nu\right) p_\infty^2
	+
	\mathcal{O}(p_\infty^4)
	\right]\\
	&
	+\mathcal{O}(G^5)\,.
	\nonumber
\end{align}
The change of the mass dipole component along the direction of motion \eqref{eq:Jy0final} is thus sensitive to the ratio of the cutoffs $\mu_\text{IR}$, $\Lambda$, but this ambiguity takes the form of a pure time translation
\begin{equation}	
	J^{\alpha\beta} + \delta{\mathcal{J}}^{\alpha\beta} \mapsto  
	J^{\alpha\beta} + \delta{\mathcal{J}}^{\alpha\beta}  + a^{[\alpha} P^{\beta]}
\end{equation}
with $a^{\alpha} = (a,0,0,0)$ in the center-of-mass frame.

Considering instead the ``subtracted'' mass dipole moment, $M_i$, which is free from the time-translation ambiguity thanks to the removal of the recoil-induced drift and thanks to the TT projection, we find, by summing  \eqref{eq:boldM1} and \eqref{eq:calJ10TT}
\begin{equation}
	\begin{split}
		{M}_{x}
		&=
		\frac{G^2 m^3}{b}
		p_\infty^3
		\nu^2	\sqrt{1-4\nu}
		\left[
		\frac{48}{35}
		+
		\left(
		\frac{568}{315}
		-
		\frac{88}{105}\,\nu
		\right)
		p_\infty^2
		+
		\mathcal{O}(p_\infty^4)
		\right]
		\\
		&+
		\frac{G^3 m^4 \pi}{b^2}
		p_\infty
		\nu^2	\sqrt{1-4\nu}
		\Bigg[
		\frac{391}{105}
		+
		\left(
		\frac{557}{140}
		-
		\frac{319}{105}\,\nu
		\right)
		p_\infty^2
		+
		\mathcal{O}(p_\infty^4)
		\Bigg]
		\\
		&+
		\frac{G^4 m^5}{b^3 p_\infty}
		\nu^2	\sqrt{1-4\nu}
		\Bigg[
		\frac{10376}{315}
		+
		\left(\frac{56068}{1575}-\frac{2420}{63}\,\nu\right) p_\infty^2
		+
		\mathcal{O}(p_\infty^4)
		\Bigg]
		\\
		&+
		\frac{G^4 m^5}{b^3}\,p_\infty^2
		\nu^2	\sqrt{1-4\nu}
		\left[
		\frac{704}{9}
		+\left(\frac{1600}{63}
		-\frac{125344}{1575}\,\nu\right) p_\infty^2
		+
		\mathcal{O}(p_\infty^4)
		\right],
		\\
		&+\mathcal{O}(G^5)\,.
	\end{split}
\end{equation}
which corresponds to the total change in the subtracted mass-dipole component along the impact parameter $b_e^\mu$.
Finally,  from \eqref{eq:boldM2}, \eqref{eq:calJ20TT} and \eqref{eq:regexpl}, we obtain
\begin{align}\label{eq:Myfinal}
	M_{y}
	&=
	\frac{G^4  \pi m^5}{b^3}\,p_\infty^2
	\nu^2	\sqrt{1-4\nu}
	\left[
	\frac{63}{10}
	+
	\left(
	-\frac{1021}{525}-\frac{203947}{75600} 
	\,\nu\right) p_\infty^2
	+
	\mathcal{O}(p_\infty^4)
	\right]\\
	&
	+\mathcal{O}(G^5)\,.
	\nonumber
\end{align}

\section{Conclusions}
\label{sec:conclusions}

In this work, we used the eikonal operator to describe the final state of a scattering between two classical massive scalar particles. This operators encodes all the classical observables relevant to the dynamics of gravitational binaries. We focused in particular on the properties of the radiation produced in the scattering recalling in Section~\ref{sec:P} the results for the radiated linear momentum and then discussing in some detail the full tensor encoding the angular momentum and the mass dipole in Section~\ref{sec:J}. This observable is sensitive to several aspects of the soft part of the radiation spectrum and the static contributions to the asymptotic gravitational field. We showed that the eikonal formalism can capture all these contributions separating in a neat way those related to the physical radiation (see Section~\ref{ssec:boldJ}) and those related to the zero-frequency modes (see Section~\ref{ssec:calJ}).

Up to 3PM, the full Lorentz-covariant results were already obtained within the eikonal formalism in~\cite{DiVecchia:2022owy,DiVecchia:2022piu} finding agreement with~\cite{Damour:2020tta,Manohar:2022dea}. It would of course be interesting to extend this analysis to 4PM for generic velocities. Here instead at $\mathcal{O}(G^4)$ we provide full PM expressions only for the static contributions, while for the radiative one we revert to the PN expansion. Summing all terms at the same PN order we find perfect agreement for the spatial components of the angular momentum tensor with~\cite{Bini:2021gat,Bini:2022enm}.

An interesting feature of the mass dipole component $J_{y0}$, where the $y$-direction is aligned with the particles' ``average'' velocity as in \eqref{eq:refvec}, is that it receives starting at 4PM a contribution from the tail terms of the waveform. This means that it displays a dependence on an arbitrary scale that can be changed by a shift of the retarded time. This issue, which appears to be a generic feature of the covariance of $J_{\alpha\beta}$ and emerges separately for the radiative and the static contributions, can be resolved by considering instead the time-translation invariant and ``TT-projected'' quantity $M_i$ defined in the center-of-mass frame. However, as verified explicitly at $\mathcal{O}(G^2)$ in the traditional approach~\cite{Bini:2022wrq} and in the eikonal formalism at $\mathcal{O}(G^3)$ \cite{DiVecchia:2022piu}, there is a balance law between the total mechanical and the gravitational angular momenta ($\Delta L_{\alpha\beta}$ and $J_{\alpha\beta} $ respectively). Indeed, letting
	\begin{equation}
		i\Delta L_{1\alpha\beta}= 
		\left(\eta_{\mu\rho}\eta_{\nu\sigma}-\tfrac{1}{D-2}\,\eta_{\mu\nu}\eta_{\rho\sigma}\right)	
		\int_k 
		\tilde{\tau}^{\ast\mu\nu}
		\Bigg(b^{\phantom{j}}_{1[\alpha} \frac{\overset{\leftrightarrow}{\partial}}{\partial b_1^{\beta]}} + \tilde{u}^{\phantom{j}}_{1[\alpha} \frac{\overset{\leftrightarrow}{\partial}}{\partial \tilde{u}_1^{\beta]}} \Bigg)
		\tilde{\tau}^{\rho\sigma}
	\end{equation}
	and similarly for $\Delta L_{2\alpha\beta}$, it follows from \eqref{eq:JmunuGaugeInv} that $\Delta L^{\alpha\beta}
	=
	\Delta L_1^{\alpha\beta}
	+
	\Delta L_2^{\alpha\beta}$ obeys
	\begin{equation}
		\Delta L_{\alpha\beta}
		=
		-
		J_{\alpha\beta}\,,
	\end{equation}
	as one can check by using the fact that $\tilde{\tau}^{\mu\nu}$ is a symmetric rank-two tensor constructed from $\tilde{u}_1^\mu$, $\tilde{u}_2^\mu$, $b_1^\mu$, $b_2^\mu$ and $k^\mu$. It was noticed~\cite{DiVecchia:2022piu} that this balance law holds separately for the static and the radiative parts at $\mathcal{O}(G^3)$ and we expect the same to be true at $\mathcal{O}(G^4)$ as well. So, the same dependence on the scale related to time shifts should appear in $\Delta L_{y0}$ both in the radiative and the static sector. It would be interesting to calculate explicitly the $\mathcal{O}(G^4)$ of $\Delta L_{\alpha\beta}$ to check that the arbitrary scale cancels and to define regularized mechanical and radiative angular momenta by moving appropriately the logarithmic terms between the two sides of the balance law. This is reminiscent of the approach taken by~\cite{Riva:2023xxm} to define a supertranslation invariant angular momentum and of course it would be very useful to extend that approach to $\mathcal{O}(G^4)$ as well.

Another interesting development is to apply to boundary to bound map~\cite{Kalin:2019rwq,Kalin:2019inp,Cho:2021arx} to the mass dipole studied here and calculate the mass dipole components for bound systems also beyond the leading PN contribution which is already discussed in~\cite{Blanchet:2013haa}.

\subsection*{Acknowledgments}
We are grateful to Francesco Alessio, Luc Blanchet, Geoffrey Comp\`ere, Poul Henrik Damgaard, Paolo Di Vecchia, Guillaume Faye, Alessandro Georgoudis, Vasco Gon\c calves, Gustav Jakobsen, Julio Parra-Martinez, Ludovic Plant\'e, Massimo Porrati, Massimiliano Riva, Ali Seraj, Chia-Hsien Shen, Gabriele Veneziano, Filippo Vernizzi for helpful discussions and communications.
C.~H. is supported by UK Research and Innovation (UKRI) under the UK government’s Horizon Europe funding guarantee [grant EP/X037312/1 ``EikoGrav: Eikonal Exponentiation and Gravitational Waves'']. R.~R. is partially supported by the UK EPSRC grant ``CFT and Gravity: Heavy States and Black Holes'' EP/W019663/1 and the STFC grants ``Amplitudes, Strings and Duality'', grant numbers ST/T000686/1 and ST/X00063X/1. No new data were generated or analysed during this study.

\appendix

\section{Time-domain and frequency-domain integrals}
\label{app:FT}

Letting
\begin{equation}
 f(t) = \int_{-\infty}^{+\infty} e^{-i \omega t}  \tilde f(\omega) \, \frac{d\omega}{2\pi}
\end{equation}
with $\tilde f(\omega)^\ast= \tilde f(-\omega)$ so that $f(t)$ is real, and similarly for $g(t)$,
we have the following identities,
\begin{subequations}
\begin{align}
\label{eq:f_g}
\int_{-\infty}^{+\infty} f(t)\, g(t)\, dt &= \int_{0}^{\infty} 
\left[
\tilde g(\omega)^\ast \tilde f(\omega)
+
\tilde f(\omega)^\ast\, \tilde g(\omega)
\right]
\frac{d\omega}{2\pi}\,,\\
\label{eq:dotf_g}
\int_{-\infty}^{+\infty} \dot f(t)\, g(t)\, dt &= \int_{0}^{\infty} 
\omega
\left[
\tilde g(\omega)^\ast \tilde f(\omega)
-
\tilde f(\omega)^\ast\, \tilde g(\omega)
\right]
\frac{d\omega}{2 i \pi}\,,\\
\label{eq:dotf_dotg}
\int_{-\infty}^{+\infty} \dot f(t)\, \dot g(t)\, dt &= \int_{0}^{\infty} 
\omega^2
\left[
\tilde g(\omega)^\ast \tilde f(\omega)
+
\tilde f(\omega)^\ast\, \tilde g(\omega)
\right]
\frac{d\omega}{2\pi}\,,\\
\label{eq:tdotf_dotf}
\int_{-\infty}^{+\infty} t  \dot f(t)^2 dt &= \int_{0}^{\infty} 
\omega
\left[
\tilde f(\omega)^\ast  \omega \tilde f'(\omega)
-
\omega \tilde f'(\omega)^\ast \tilde f(\omega)
\right]
\frac{d\omega}{2 i \pi}\,,
\end{align}
\end{subequations}
where a superscript dot and a prime denote derivatives with respect to time and to frequency. Note that, for constant $g(t)=c_0$, Eq.~\eqref{eq:dotf_g} gives
\begin{equation}
	f(t=+\infty) - f(t=-\infty) = -i \lim_{\omega\to0^+} \omega \left[\tilde f(\omega)-\tilde f(\omega)^\ast\right].
\end{equation}

\section{Uplifting the TT formula}
\label{app:upliftTT}
In this appendix, we provide the general uplift of the TT formula for the angular momentum to its covariant counterpart, thus extending the derivation given in \cite{DiVecchia:2022owy}. We introduce the shorthand notation
\begin{equation}
	d_{\alpha\beta} = k^{\phantom{\beta}}_{[\alpha}\frac{\partial}{\partial k^{\beta]}}
\end{equation}
for the differential operator that preserves the mass-shell constraint $k^2=0$.

\subsection{Spin one}
We begin from the spin-one case by considering a generic $F^\mu(k)$ such that
\begin{equation}
	S(k) = k_\mu F^\mu(k)
\end{equation}
and its transverse projection,
\begin{equation}
	f_\mu = \Pi_{\mu\nu} F^\nu
\end{equation}
with $\Pi_{\mu\nu}$ as defined in \eqref{eq:Tprojector} in terms of a reference vector $\lambda^\mu$ obeying $\lambda^2=0$, $\lambda\cdot k=-1$.
Using the identities,
\begin{equation}\label{eq:lambdakproperties}
	\lambda^\mu d_{\alpha\beta} \lambda_{\mu} = 0\,,\qquad
	k^\mu d_{\alpha\beta} \lambda_{\mu} + k_{[\alpha}\lambda_{\beta]}=0\,,
\end{equation}
which follow uniquely from the above properties obeyed by $\lambda^\mu$, and integrating by parts, one can show that the ``transverse'' formula
\begin{equation}
	i\mathcal{J}^\text{T}_{\alpha\beta}
	=
	\int_k \left(
	f^\ast_\mu \overset{\leftrightarrow}{d}_{\alpha\beta} f^{\mu}
	+
	f^\ast_{[\alpha}
	f^{\phantom{\ast}}_{\beta]}
	\right)
\end{equation}
is equivalent to
\begin{equation}\label{eq:upliftSpin1}
	i\mathcal{J}^\text{T}_{\alpha\beta}
	=
	i\mathcal{J}^\text{vec}_{\alpha\beta}
	-
	\int_k \left[
	\lambda\cdot F\,
	d_{\alpha\beta}S^\ast
	+
	S^\ast \left(
	F^\mu d_{\alpha\beta} \lambda_{\mu}
	+
	F_{[\alpha} \lambda_{\beta]}
	\right)
	-
	\text{c.c.}
	\right],
\end{equation}
where the first term on the right-hand side provides the manifestly $\lambda$-independent uplift, 
\begin{equation}
	i\mathcal{J}^\text{vec}_{\alpha\beta}
	=
	\int_k \left(
	F^\ast_\mu \overset{\leftrightarrow}{d}_{\alpha\beta} F^{\mu}
	+
	F^\ast_{[\alpha}
	F^{\phantom{\ast}}_{\beta]}
	\right),
\end{equation}
and the remaining terms on the right-hand side of \eqref{eq:upliftSpin1} vanish if $S=k_\mu F^\mu$ is identically zero. 

Eq.~\eqref{eq:upliftSpin1} holds for any $\lambda^\mu$ obeying $\lambda^2=0$, $\lambda\cdot k=-1$ and for any $F^\mu$. Specializing to the parametrization $k^\mu = \omega(1,\hat n)$ and $\lambda^\mu = \frac{1}{2\omega}(1,-\hat n)$ as in \eqref{eq:ndecomp}, \eqref{eq:lambdadecomp} in a given frame, one can show that, for any $T^\mu$,
\begin{equation}\label{eq:lambdaspecialproperties}
	T^\mu d_{ij} \lambda_{\mu} + T_{[i} \lambda_{j]} = 0\,,
	\qquad
	T^\mu d_{i0} \lambda_{\mu} + T_{[i} \lambda_{0]} = - \frac{1}{\omega}\, \Pi_{i\mu}T^{\mu}\,. 
\end{equation}
Further considering the explicit soft photon factor, one has (like for gravity, a factor of $\frac12$ is introduced in $S$ below to avoid double counting)
\begin{equation}
	F^\mu = \sum_a \frac{\eta_a\,e_a\, p_a^\mu}{p_a\cdot k-i0}\,,\qquad
	\frac{1}{\omega}\, S = -i\pi \mathsf{Q}\,\delta(\omega)\,,\qquad
	\mathsf{Q} = \sum_{a\in\text{in}}e_a
	=
	\sum_{a\in\text{out}} e_a
\end{equation}
with $\mathsf{Q}$ the total Coulombic charge. Then, by applying  \eqref{eq:upliftSpin1} in the center-of-mass frame, one obtains
\begin{equation}\label{eq:calJTvec}
	\mathcal{J}^\text{T}_{ij}=\mathcal{J}^\text{vec}_{ij}\,,
	\qquad
	\mathcal{J}^{\text{T}}_{i0}
	=
	\mathcal{J}^{\text{vec}}_{i0}
	+
	\frac{\mathsf{Q}}{4\pi}\, \sum_{a} \frac{e_a}{\sigma^2_a-1} \left(
	\frac{\operatorname{arccosh}\sigma_a}{\sqrt{\sigma^2_a-1}}
	-
	\sigma_a
	\right)
	\frac{p_a^i}{m_a}
\end{equation}
in terms of the variables $\sigma_a = E_a/m_a$ (with $E_a$ the center-of-mass energy of the state $a$). Combining the expressions in \eqref{eq:calJTvec} with \cite{DiVecchia:2022owy}
\begin{equation}
	\mathcal{J}^\text{vec}_{\alpha\beta}
	=
	-
	\frac{1}{16\pi}
	\sum_{a,b} \frac{e_a e_b}{\sigma^2_{ab}-1} \left(
	\frac{\operatorname{arccosh}\sigma_{ab}}{\sqrt{\sigma^2_{ab}-1}}
	-
	\sigma_{ab}
	\right)
	(\eta_a-\eta_b)\,
	\frac{
		p_a^{[\alpha}p_b^{\beta]}
	}{m_am_b}
\end{equation}
gives the full result for $\mathcal{J}_{\alpha\beta}^\text{T}$ in the center-of-mass frame.

\subsection{Spin two}
Let us move to the spin-two case relevant for gravity. We thus consider a generic symmetric $F^{\mu\nu}(k)$, for which we define
\begin{equation}
	R^\mu(k) = k_\nu F^{\mu\nu}(k)\,,\qquad
	R(k) = k_\mu R^{\mu}(k)\,,
\end{equation}
and its TT projection
\begin{equation}
	f^{\mu\nu} = \Pi_{\mu\nu,\rho\sigma} F^{\rho\sigma}\,,
\end{equation}
where $\Pi_{\mu\nu,\rho\sigma}$ is defined in \eqref{eq:TTprojector}.
Starting from the TT formula
\begin{equation}
	i \mathcal{J}^\text{TT}_{\alpha\beta}
	=
	\int_k 
	\left(
	f^\ast_{\mu\nu} \overset{\leftrightarrow}{d}_{\alpha\beta}
	f^{\mu\nu} + 2 f^\ast_{\mu[\alpha} f_{\beta]}^\mu
	\right),
\end{equation}
one can again use the properties \eqref{eq:lambdakproperties} and integrate by parts to show that
\begin{equation}\label{eq:upliftSpin2}
	\begin{split}
		i \mathcal{J}^\text{TT}_{\alpha\beta}
		=
		i \mathcal{J}_{\alpha\beta}
		&+
		\int_k 
		\Big[	
		\tfrac{D-4}{D-2}\,R^\ast\cdot \lambda\, d_{\alpha\beta} \left(R\cdot \lambda\right)	
		+
		2R\cdot \lambda 
		\left(
		R^{\ast\mu} d_{\alpha\beta} \lambda_\mu + R^\ast_{[\alpha} \lambda^{\phantom{\ast}}_{\beta]}
		\right)
		\\
		&
		-
		\lambda\cdot F\cdot \lambda \, d_{\alpha\beta}  R^\ast
		-
		2R^\ast 
		\left(
		\lambda\cdot F^\mu d_{\alpha\beta} \lambda_{\mu}
		+
		\lambda\cdot F_{[\alpha}\lambda_{\beta]}
		\right)
		\\
		&
		-
		2 \lambda\cdot F^\mu \, d_{\alpha\beta} R^\ast_{\mu}
		-
		2 R^\ast_\mu 
		\left(
		F^{\mu\nu}d_{\alpha\beta}\lambda_\nu + F^\mu_{[\alpha} \lambda^{\phantom{\mu}}_{\beta]}
		\right)
		\\
		&
		+
		\tfrac{2}{D-2}\, \eta^{\mu\nu}F_{\mu\nu} \, d_{\alpha\beta}(R^\ast\cdot \lambda)
		-
		2\lambda\cdot F_{[\alpha}R^\ast_{\beta]}
		-
		\text{c.c.}
		\Big],
	\end{split}
\end{equation}
where
\begin{equation}
	i \mathcal{J}_{\alpha\beta}
	=
	\int_k 
	\left[
	\left(\eta_{\mu\rho}\eta_{\nu\sigma}
	-
	\tfrac{1}{D-2}
	\,\eta_{\mu\nu}\eta_{\rho\sigma}
	\right)
	F^\ast_{\mu\nu} \overset{\leftrightarrow}{d}_{\alpha\beta}
	F^{\mu\nu} + 2 F^\ast_{\mu[\alpha} F_{\beta]}^\mu
	\right]
\end{equation}
is the manifestly $\lambda$-independent uplift, and the remaining terms on the right-hand side of \eqref{eq:upliftSpin2} vanish if $R^\mu = k_{\nu} F^{\mu\nu}$ is identically zero. 
In particular, \eqref{eq:upliftSpin2} ensures the equivalence between \eqref{eq:JmunuTT} and \eqref{eq:JmunuGaugeInvw}, thanks to the \emph{exact} transversality property \eqref{eq:upliftW}.

Eq.~\eqref{eq:upliftSpin2} holds for any $\lambda^\mu$ satisfying $\lambda^2=0$, $\lambda\cdot k=-1$ and for generic $F^{\mu\nu}$. One can then specialize to the choice $k^\mu = \omega(1,\hat n)$ and  $\lambda^\mu = \frac{1}{2\omega}(1,-\hat n)$ as in \eqref{eq:lambdadecomp}, for which one has the additional relations \eqref{eq:lambdaspecialproperties}, and apply \eqref{eq:upliftSpin2} to the soft factor \eqref{eq:Fmunu} (or \eqref{eq:FmunuT}), in which case $\frac{1}{\omega}\,R^\mu=n_\nu F^{\mu\nu}$ is given by \eqref{eq:nontransv}. In this way, one obtains the results presented in the main body of the text for static effects in the gravitational emission of angular momentum, in particular \eqref{eq:calJalphabetaFinal} and hence \eqref{eq:deltacalJTT}, in the center-of-mass frame. 

One can also apply \eqref{eq:upliftSpin2} in a generic frame. Letting a prime denote quantities evaluated in that frame, and choosing $k'^\mu = \omega'(1,\hat n')$ and reference vector $\lambda'^\mu = \frac{1}{2\omega'}(1,-\hat n')$, one obtains in particular
\begin{equation}\label{eq:generalTTij}
	\mathcal{J}^{\text{TT}\, \prime}_{ij}=\mathcal{J}^{\prime}_{ij}-2G\sum_a c\left(\sigma'_a\right) \sum_{b\in\text{in}} p_b^{\prime\,[i} p_a^{\prime\,j]}
\end{equation}
with $\sigma'_a = E'_a/m_a$. Note that, of course, \eqref{eq:generalTTij} is in general different from the result obtained by evaluating \eqref{eq:calJalphabetaFinal} in the primed frame, owing to the two different choices of reference vectors (while the two formulas do agree in the center-of-mass frame).
From \eqref{eq:generalTTij}, choosing the frame where particle 1 is initially at rest, $p_1^{\prime\,i}=0$, one recovers $\mathcal{J}^{\text{TT}\,\prime}_{xy}=GQ(b)m_2 p_\infty \mathcal{I}(\sigma) + \mathcal{O}(G^3)$, in agreement with \cite{Jakobsen:2021smu,Mougiakakos:2021ckm}, while $\mathcal{J}^{\prime}_{xy}=\frac{1}{2}\,GQ(b)m_2 p_\infty \mathcal{I}(\sigma) + \mathcal{O}(G^3)$ as in \cite{Manohar:2022dea,DiVecchia:2022owy}.

\providecommand{\href}[2]{#2}\begingroup\raggedright\endgroup

\end{document}